%% file: ms.tex
\documentclass[12pt,preprint]{aastex}

\shorttitle{NGC 6569}
\shortauthors{Johnson et al.}

\newcommand\iso[2]{$^{\rm #1}$#2}

\begin{document}

\title{Exploring the Chemical Composition and Double Horizontal Branch of the
Bulge Globular Cluster NGC 6569}

\author{
Christian I. Johnson\altaffilmark{1,2},
R. Michael Rich\altaffilmark{3},
Nelson Caldwell\altaffilmark{1},
Mario Mateo\altaffilmark{4}, 
John I. Bailey, III\altaffilmark{5}, 
Edward W. Olszewski\altaffilmark{6}, and
Matthew G. Walker\altaffilmark{7}
}

\altaffiltext{1}{Harvard--Smithsonian Center for Astrophysics, 60 Garden
Street, MS--15, Cambridge, MA 02138, USA; cjohnson@cfa.harvard.edu;
ncaldwell@cfa.harvard.edu}

\altaffiltext{2}{Clay Fellow}

\altaffiltext{3}{Department of Physics and Astronomy, UCLA, 430 Portola Plaza,
Box 951547, Los Angeles, CA 90095-1547, USA; rmr@astro.ucla.edu}

\altaffiltext{4}{Department of Astronomy, University of Michigan, Ann Arbor,
MI 48109, USA; mmateo@umich.edu}

\altaffiltext{5}{Leiden Observatory, Leiden University, P. O. Box 9513, 2300RA
Leiden, The Netherlands; baileyji@strw.leidenuniv.nl}

\altaffiltext{6}{Steward Observatory, The University of Arizona, 933 N. Cherry
Avenue, Tucson, AZ 85721, USA; eolszewski@as.arizona.edu}

\altaffiltext{7}{McWilliams Center for Cosmology, Department of Physics,
Carnegie Mellon University, 5000 Forbes Avenue, Pittsburgh, PA 15213, USA;
mgwalker@andrew.cmu.edu}

\begin{abstract}

Photometric and spectroscopic analyses have shown that the Galactic bulge 
cluster Terzan 5 hosts several populations with different metallicities and 
ages that manifest as a double red horizontal branch (HB).  A recent
investigation of the massive bulge cluster NGC 6569 revealed a similar, though
less extended, HB luminosity split, but little is known about the cluster's
detailed chemical composition.  Therefore, we have used high resolution spectra
from the Magellan--M2FS and VLT--FLAMES spectrographs to investigate the 
chemical compositions and radial velocity distributions of red giant branch and
HB stars in NGC 6569.  We found the cluster to have a mean heliocentric radial 
velocity of --48.8 km s$^{\rm -1}$ ($\sigma$ = 5.3 km s$^{\rm -1}$; 148 stars) 
and $\langle$[Fe/H]$\rangle$ = --0.87 dex (19 stars), but the cluster's 0.05 
dex [Fe/H] dispersion precludes a significant metallicity spread.  NGC 6569 
exhibits light and heavy element distributions that are common among old 
bulge/inner Galaxy globular clusters, including clear (anti--)correlations 
between [O/Fe], [Na/Fe], and [Al/Fe].  The light element data suggest NGC 6569 
may be composed of at least two distinct populations, and the cluster's low 
$\langle$[La/Eu]$\rangle$ = --0.11 dex indicates significant pollution with 
r--process material.  We confirm that both HBs contain cluster members, but 
metallicity and light element variations are largely ruled out as sources for 
the luminosity difference.  However, He mass fraction differences as small as 
$\Delta$Y $\sim$ 0.02 cannot be ruled out and may be sufficient to reproduce 
the double HB.

\end{abstract}


\section{INTRODUCTION}

Correlated star--to--star abundance variations involving elements ranging from 
at least carbon to aluminum are common within nearly all old globular clusters 
\citep[see reviews by][]{Kraft94,Gratton04,Gratton12}, and are driven by both 
evolutionary processes \citep[e.g., first dredge--up and extra 
mixing;][]{Charbonnel95,Denissenkov03,DAntona07} and pollution from previous
generations of more massive stars \citep[e.g.,][]{Decressin07,deMink09,
Bastian13,Ventura13,Denissenkov14}.  Although the light element
abundances can vary by more than a factor of 10 within a single cluster, for 
most systems the heavier $\alpha$ and Fe--peak element abundance dispersions
are generally $\la$ 0.05 dex \citep[e.g.,][]{Sneden04,Carretta09a,Carretta10a}.
For the neutron--capture elements, a small number of clusters exhibit 
significant ($\ga$ 0.3 dex) abundance dispersions that may be attributed to
primordial enrichment via the rapid neutron--capture process 
\citep[r--process;][]{Roederer11}, but in most cases the intrinsic heavy
element [X/Fe]\footnote{[A/B] 
$\equiv$ log(N$_{\rm A}$/N$_{\rm B}$)$_{\rm star}$ --
log(N$_{\rm A}$/N$_{\rm B}$)$_{\sun}$ and log $\epsilon$(A) $\equiv$
log(N$_{\rm A}$/N$_{\rm H}$) + 12.0 for elements A and B.} variations do not 
exceed $\sim$0.2 dex \citep[e.g.,][]{DOrazi10}.  Furthermore, a majority of 
clusters have [Ba/Eu] and [La/Eu] ratios that are consistent with an 
r--process dominated composition \citep[e.g.,][]{Gratton04}.  Taken together, 
the mean composition characteristics outlined above suggest that old globular 
clusters formed rapidly ($\la$ 100 Myr), self--enriched with the products of 
proton--capture nucleosynthesis, did not generally retain the ejecta of core 
collapse supernovae, and ceased star formation before the winds of $\la$ 4 
M$_{\rm \sun}$ asymptotic giant branch (AGB) stars could pollute cluster 
interstellar mediums with the products of slow neutron--capture process 
(s--process) nucleosynthesis.

However, new evidence indicates that several of the most massive Galactic 
globular clusters contain stellar populations with different light \emph{and} 
heavy element abundances.  These ``iron--complex" clusters exhibit significant 
[Fe/H] dispersions and share a common trait that metallicity and s--process
enhancements are strongly correlated \citep[e.g.,][]{Norris95,Smith00,Johnson10,
Marino11a,Marino11b,Yong14,Johnson15a,Marino15,DaCosta16a,
Johnson17a}\footnote{Although the s--process abundance variations are generally
reproduced by all analyses, the extent of the [Fe/H] variations have been 
questioned for several potential iron--complex clusters, including M 22 
\citep[][but see also Lee et al. 2015]{Mucciarelli15}, M 2 \citep{Lardo16}, and
NGC 1851 \citep{Villanova10}.  Additionally, the two most massive iron--complex
clusters $\omega$ Cen and M 54 share similar metallicity and light element 
distributions \citep{Carretta10b}, but M 54's s--process abundances have not 
been extensively explored.}.  Since second peak s--process elements are thought
to be produced during the late stage evolution of low and intermediate mass AGB
stars \citep[e.g.,][]{Busso99}, the correlation between [Ba,La/Fe] and [Fe/H] 
is consistent with the idea that iron--complex clusters experienced prolonged 
star formation and chemical enrichment compared to monometallic systems.  
Although the origin of iron--complex clusters is not yet clear, the strong 
retrograde orbit exhibited by $\omega$ Cen \citep{Dinescu99,Tsuchiya03} 
combined with M 54's likely origin in the Sagittarius dwarf spheroidal galaxy 
\citep[e.g.,][]{Bellazzini08} makes plausible the idea that at least some 
iron--complex clusters may have been accreted by the Milky Way \citep{Bekki16,
DaCosta16a}.  In such a scenario, iron--complex clusters would represent the 
remnants of minor merger events with the Galaxy, and together with their 
progenitor systems would have contributed as ``building blocks" to the 
formation of the halo, disk, and bulge.

In this context, the Galactic bulge globular cluster Terzan 5 is particularly 
interesting.  Similar to more conventional iron--complex clusters,  Terzan 5 
exhibits a large metallicity spread with distinct populations near [Fe/H] 
$\approx$ --0.8, --0.25, and $+$0.3 dex \citep{Ferraro09,Origlia11,Origlia13,
Massari14}\footnote{A confirmation of similar s--process element abundance
variations is not yet available because the cluster is obscured by high
foreground reddening \citep[E(B--V) $>$ 2;][]{Massari12}.}.  On the other hand,
Terzan 5 may be an entirely different class of object.  For example, 
\citet{Massari15} have ruled out an accretion origin for Terzan 5 based on 
proper motion measurements, and the cluster's metallicity dispersion appears to
be correlated with an approximately 7.5 Gyr age spread \citep{Ferraro16}.  A 
comparison of the cluster's [Fe/H], age, [$\alpha$/Fe], and light element 
abundance distributions with those of bulge field stars shows remarkable 
similarities, and raises the possibility that Terzan 5 may be a remnant 
primordial building block of the bulge rather than a genuine globular cluster 
\citep{Ferraro09,Origlia11,Ferraro16}.  However, a recent analysis by 
\citet{Schiavon17a} found some evidence of (anti--)correlations between C, N, 
Na, and Al in a small sample of stars so the possibility remains that Terzan 5 
may be an iron--complex cluster.

One of the interesting aspects about Terzan 5 is that it was designated for 
spectroscopic follow--up because \citet{Ferraro09} discovered the presence of 
two red horizontal branches (HBs) separated in the K--band by 0.3 magnitudes.
The double HB feature in Terzan 5 has been linked to the cluster's metallicity
and age spread \citep{Ferraro09,Ferraro16}.  A similar analysis by 
\citet{Mauro12} discovered that the two metal--rich bulge clusters NGC 6440 
and NGC 6569 also exhibit double red HBs.  However, in both of those cases the 
HBs were separated by only $\sim$0.1 mag in K$_{\rm S}$, and the authors were 
not able to determine the cause of the HB luminosity differences.  Recently, 
\citet{Munoz17} examined the chemical composition of RGB stars in NGC 6440, 
but did not find evidence of an internal [Fe/H] spread nor any peculiar light 
element abundances that would explain the double HB feature.  \citet{Munoz17} 
did find that NGC 6440 exhibits large ranges in [Na/Fe] and [Al/Fe], but 
curiously did not find a significant O--Na anti--correlation.

From a chemical perspective, little is known about NGC 6569, which is an old
bulge/inner Galaxy (l,b) = ($+$0.48$\degr$, --6.68$\degr$) globular cluster that
resides approximately 3 kpc from the Galactic center 
\citep[][2010 version]{Harris96}.  \citet{Valenti11} represents the only high 
resolution spectroscopic analysis of the cluster, and did not find any evidence
supporting an intrinsic metallicity spread nor extreme light element abundance 
variations.  However, their sample size was only 6 stars and did not include 
any heavy s--process elements that would have helped identify NGC 6569 as a 
possible iron--complex cluster.  Therefore, we provide here a detailed analysis
of the cluster's chemical composition, including an examination of the 
``bright" and ``faint" HB stars, in order to gain insight into both the 
cluster's chemical enrichment history and any possible explanations for its HB 
morphology.

\section{OBSERVATIONS AND DATA REDUCTION}

\subsection{Magellan Spectroscopic Sample}

The primary data obtained for this project were acquired on 2014 June 03 for
the RGB sample and between 2016 June 28 and 2016 July 01 for the HB sample
(see Table 1).  All observations utilized the Magellan--Clay 6.5m telescope at 
Las Campanas Observatory instrumented with the Michigan--Magellan Fiber System 
\citep[M2FS; ][]{Mateo12} and MSpec spectrograph.  Although the RGB stars were
observed during excellent observing conditions (seeing $<$ 1$\arcsec$), the
HB observations were obtained in generally poor sky conditions with seeing $>$ 
1.5$\arcsec$.  

\begin{figure}
\epsscale{1.00}
\plotone{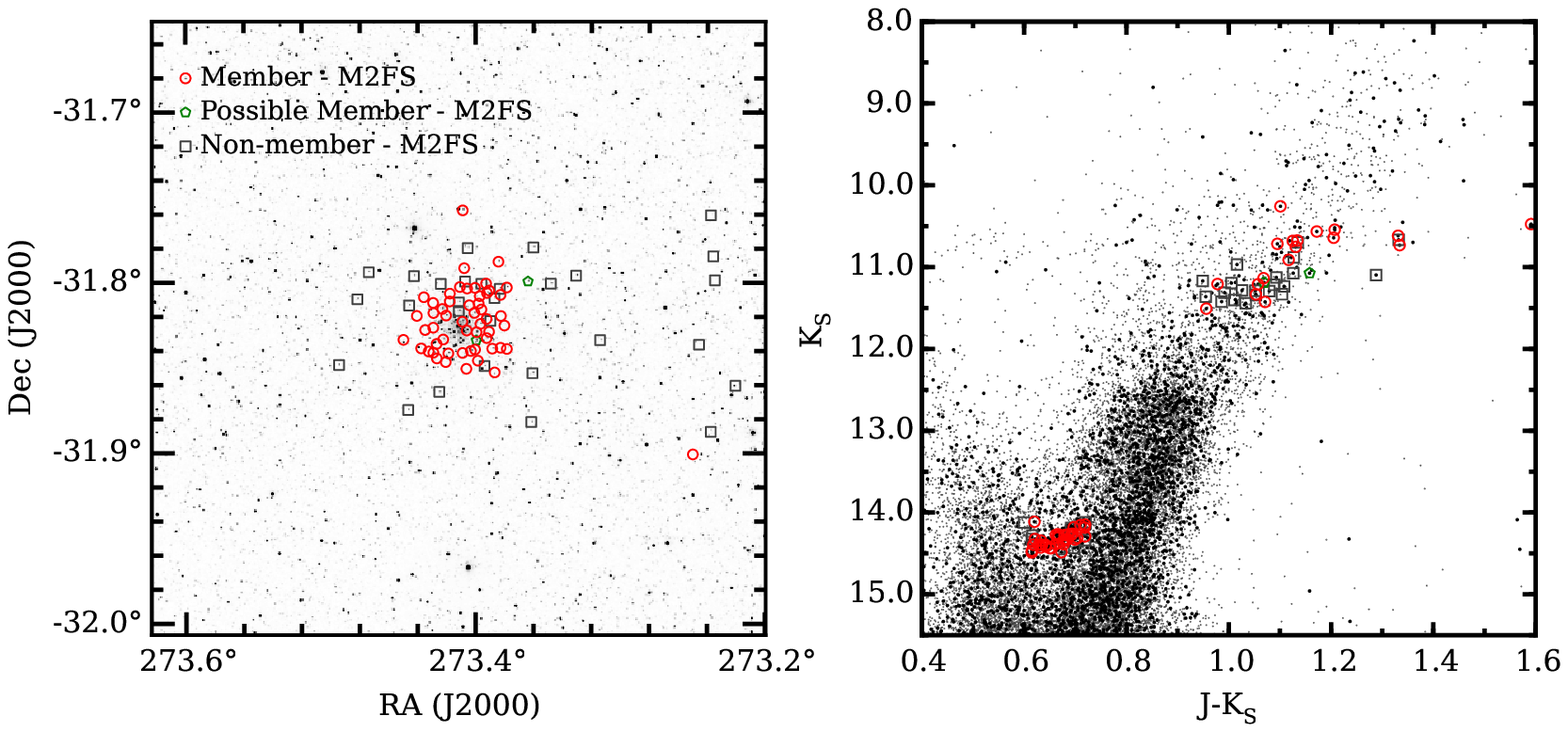}
\caption{Left: the sky coordinates of all member (open red circles), possible
member (open green pentagons), and non--member (open grey boxes) stars observed
with M2FS are overplotted on a 2MASS J--band image centered on NGC 6569.  The
stars designated as possible members have radial velocities consistent with
cluster membership but lack [Fe/H] measurements.  Right: the member, possible
member, and non--member stars observed with M2FS are plotted on a K$_{\rm S}$ 
versus J--K$_{\rm S}$ color--magnitude diagram using data from the VVV survey. 
For stars where VVV J and/or K$_{\rm S}$ photometry was unavailable, we have 
substituted 2MASS magnitudes.  Note that the few member stars with 
J--K$_{\rm S}$ $>$  1.3 either have large magnitude errors or exhibit 
significant differences between their VVV and 2MASS magnitudes.  The filled 
black circles indicate all stars within 5$\arcmin$ of the cluster center, and 
the open grey circles indicate stars with projected radial distances between 
5--15$\arcmin$ from the cluster center.  The remaining colors and symbols are 
the same as in the left panel.}
\label{f1}
\end{figure}

Target coordinates and photometry for the RGB sample were taken from the Two 
Micron All Sky Survey \citep[2MASS; ][]{Skrutskie06}.  A selection function was
generated by identifying the fiducial RGB sequence in a K$_{\rm S}$ versus 
J--K$_{\rm S}$ color--magnitude diagram that included only stars within 
1$\arcmin$ of the cluster center.  The selection box was then extended to 
include stars out to $\sim$5$\arcmin$ from the cluster center, but only 
isolated (i.e., not blended in 2MASS J--band images) stars within $\sim$2 
magnitudes of the RGB--tip were drilled into the M2FS plate as potential 
targets.  Stars closer to the cluster core were also given a higher priority
ranking in an effort to mitigate bulge field star contamination.  The sky 
coordinates and photometry of all 42 RGB targets observed with M2FS are shown 
in Figure \ref{f1} and are also provided in Table 2.

For the M2FS HB sample, the target coordinates and photometry were obtained 
from the VISTA variables in the V{\'i}a L{\'a}ctea \citep[VVV; ][]{Saito12} 
survey DR1 catalog.  We used the selection boxes provided by 
Figure 3 of \citet{Mauro12} to identify targets in the ``bright" (HB--A) and 
``faint" (HB--B) HB populations.  The observed sample shown in Figure \ref{f1} 
and listed in Table 2 includes 9 HB--A stars, 17 HB--B stars, and 17 additional 
stars that are near, but just outside, the HB--A and HB--B photometric 
boundaries of \citet{Mauro12}.  Similar to the RGB sample, HB targets located 
closer to the cluster center were given a higher priority during the plate 
design process, but $\sim$15$\%$ of our sample fell between 5--10$\arcmin$ from
the cluster core.  A full list of the 2MASS and VVV star names, coordinates, 
and photometry for the HB targets is provided in Table 2.

The RGB and HB samples both utilized the same ``Bulge$\_$GC1" spectrograph
configuration described in \citet{Johnson15b} that provides continuous 
wavelength coverage from $\sim$6140--6720 \AA\ for up to 48 targets.  
Additionally, all M2FS observations were obtained with 1 $\times$ 2 (dispersion
$\times$ spatial) CCD binning, a four amplifier slow readout mode, and the 
125$\mu$m slits.  Combined with the 1.2$\arcsec$ fibers and echelle gratings,
we achieved a typical resolving power R $\equiv$ $\lambda$/$\Delta$$\lambda$ 
$\approx$ 27,000 for all observations.  

Data reduction for the RGB and HB data sets was carried out following the 
methods outlined in \citet{Johnson15b}.  Specifically, we used standard 
IRAF\footnote{IRAF is distributed by the National Optical Astronomy 
Observatory, which is operated by the Association of Universities for Research 
in Astronomy, Inc., under cooperative agreement with the National Science 
Foundation.} routines to subtract the overscan and bias levels, trim the 
overscan regions, and correct for dark current effects on each individual
amplifier image.  The \emph{imtranspose} and \emph{imjoin} IRAF tasks were then
used to rotate, align, and join the individual amplifier images into single
full CCD images.  Additional data reduction tasks, including aperture tracing,
flat--field normalization, scattered light removal, ThAr wavelength 
calibration, cosmic ray removal, fiber--to--fiber throughput correction, and 
spectrum extraction, were completed using the IRAF \emph{dohydra} routine.  The
sky fibers were extracted separately and used to create master (combined) sky 
spectra that were then subtracted from each exposure.  For the high 
signal--to--noise (S/N) ratio RGB data (S/N $>$ 50 per reduced pixel), the 
individual exposures were continuum normalized and then median combined; 
however, the HB data only had S/N $\sim$ 5--10 per reduced pixel per exposure 
so the extracted spectra were co--added before continuum normalization.  The 
final combined RGB and HB spectra had typical S/N ratios of $\sim$100 and 20
per reduced pixel, respectively.

\subsection{Very Large Telescope Spectroscopic Sample}

Given the possibility that the double HB discovered by \citet{Mauro12} could
be driven by metallicity and/or light element abundance variations, we extended
our sample by downloading FLAMES--GIRAFFE data from the ESO 
archive\footnote{Based on data obtained from the ESO Science Archive Facility 
under request number 281576.  The original data were taken as a part of 
programs 093.D--0286(A) and 193.D--0232(F).}.  The archival data included 
observations for a combination of RGB and HB stars with the HR13 and HR21 
configurations, respectively.  As is summarized in Table 1, the HR13 
VLT--FLAMES data were obtained between 2014 July 03 and 2014 August 01 and the 
HR21 data were obtained between 2015 June 22 and 2015 July 26.  Both data sets 
were binned 1 $\times$ 1 with the HR13 and HR21 spanning 6115--6395 \AA\ and 
8482--8982 \AA\ at R $\sim$ 26,400 and 18,000, respectively.

\begin{figure}
\epsscale{1.00}
\plotone{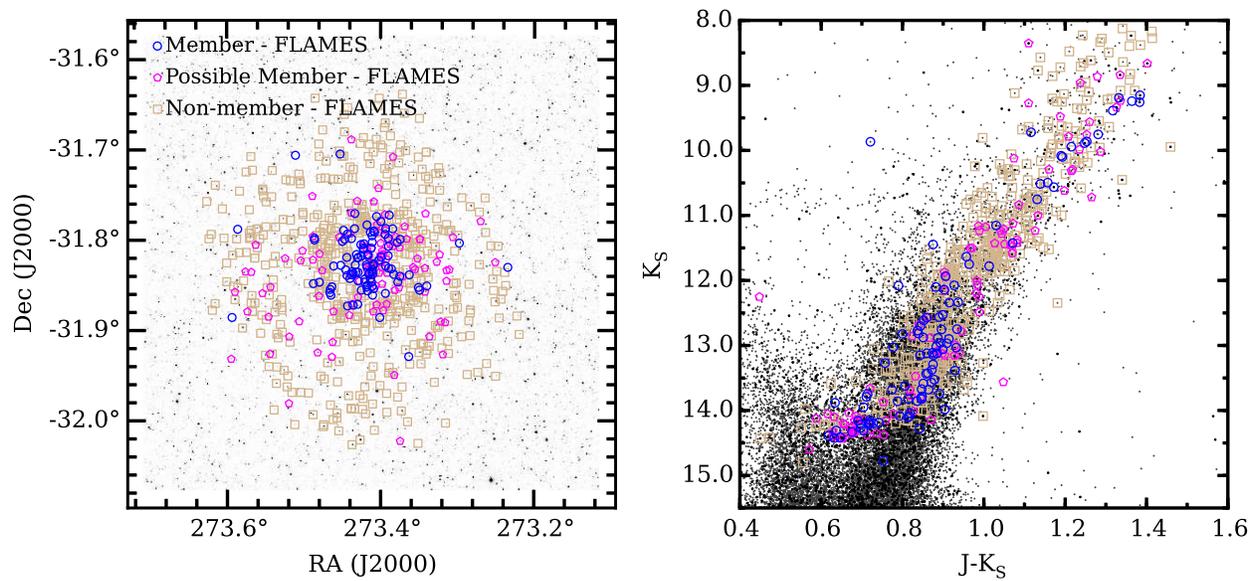}
\caption{Similar to Figure \ref{f1}, the left and right panels show the 
sky coordinates and VVV colors and magnitudes for NGC 6569 member (open
blue circles), possible member (open magenta pentagons), and non--member (open 
tan boxes) stars observed with FLAMES.}
\label{f2}
\end{figure}

Figure \ref{f2} shows that the FLAMES observations span a luminosity range that
is similar to the M2FS data, but include additional stars between the upper RGB
and HB.  When combined, the two FLAMES archival data sets provided spectra for 
$\sim$800 unique RGB and HB stars ranging from $\sim$0.05$\arcmin$ to 
12.5$\arcmin$ from the cluster center (see Figure \ref{f2}).  However, only
$\sim$115 stars were observed in the HR13 setup because the two observing runs
targeted the same stars.  Since the star names and coordinates provided in the 
FLAMES image headers did not exactly match 2MASS nor VVV, we selected the 
closest match within a radius of 2$\arcsec$ from each survey for each observed 
target.  In cases where a clear match could not be found, we retained the 
objects for radial velocity measurements but did not measure the chemical 
compositions of these stars.  A summary of the star identifiers, coordinates, 
and photometry for the FLAMES data sets is provided in Table 3.

A majority of the FLAMES data reduction was carried out using the GIRAFFE 
Base--Line Data Reduction Software (girBLDRS) package\footnote{The girBLDRS
software can be downloaded at: http://girbldrs.sourceforge.net/.}.  Similar
to the IRAF reduction of our M2FS data, we used girBLDRS to overscan and bias
correct the images, trim the overscan regions, fit and trace the apertures,
subtract scattered light, apply the flat--field corrections, fit and apply
the ThAr wavelength calibrations, and extract the object and sky spectra.
Additional data reduction procedures, including sky subtraction, spectrum
combining, continuum normalization, and telluric removal, were performed using 
standard IRAF routines.

Since the HR13 observations targeted the same stars twice, we co--added their
spectra to achieve S/N ratios of $\sim$70--100 per reduced pixel for the 
brightest stars and $\sim$30 for the HB stars.  However, since the S/N,
spectral coverage, and membership fractions (see Section 3) are considerably 
smaller in the FLAMES sample compared to the M2FS sample, we did not perform 
detailed chemical composition analyses on the HR13 data.  For the HR21 
observations, the S/N ratios from individual exposures generally exceeded
100 per reduced pixel.  As a consequence of the high S/N and resolution, we 
were able to perform radial velocity and Calcium Triplet (CaT) metallicity 
measurements on individual exposures.  We note that two of the HR21 observation
sets also targeted the same stars, but rather than combine the spectra we 
analyzed each set separately in order to estimate the measurement errors.

\section{RADIAL VELOCITIES AND CLUSTER MEMBERSHIP}

Radial velocities for all stars were measured using the XCSAO cross correlation
code \citep{Kurtz98}.  Since the M2FS and FLAMES data sets contain a 
combination of RGB and HB stars, we generated synthetic spectra representative 
of the parameter space spanned by the observations to serve as cross correlation
templates.  The fitting template returning the highest cross correlation
coefficient was then used to calculate the final velocity for each star.

For the M2FS data, we independently measured radial velocities for each order
of each exposure per star, and the mean velocities calculated with this method 
are listed in Table 2.  In all cases, a heliocentric correction was determined
for each exposure using the IRAF \emph{rvcor} routine, and the correction was
applied before averaging the individual measurements.  Similarly, the standard 
deviations of all heliocentric corrected velocity measurements for each star 
are provided as the velocity error column in Table 2.  For the 
FLAMES data, the HR13 observations and 2/7 HR21 observations targeted the same 
stars on two separate nights, and for those cases the velocity and error 
columns of Table 3 represent the mean velocities and standard deviations over
the two nights.  For the remaining HR21 fields that were only observed once, 
the velocity and error columns represent the values returned by the XCSAO code.
The heliocentric corrections for all FLAMES data were obtained from the image
headers.

In addition to stars being observed twice in the HR13 configuration and 2/7 
HR21 setups, a small number of stars were also observed more than once in
combinations of the M2FS, HR13, and HR21 fields.  The M2FS/HR13, M2FS/HR21, and
HR13/HR21 setups have 3, 21, and 4 stars in common and exhibit mean 
velocity differences of 0.0 km s$^{\rm -1}$ ($\sigma$ = 0.6 km s$^{\rm -1}$),
0.7 km s$^{\rm -1}$ ($\sigma$ = 1.0 km s$^{\rm -1}$), and 0.8 km s$^{\rm -1}$ 
($\sigma$ = 0.8 km s$^{\rm -1}$), respectively.  The mean velocity differences
between setups are comparable to the mean M2FS, HR13, and HR21 measurement 
errors for individual stars of 0.4 km s$^{\rm -1}$, 0.4 km s$^{\rm -1}$, and 
0.8 km s$^{\rm -1}$, respectively.

\begin{figure}
\epsscale{1.00}
\plotone{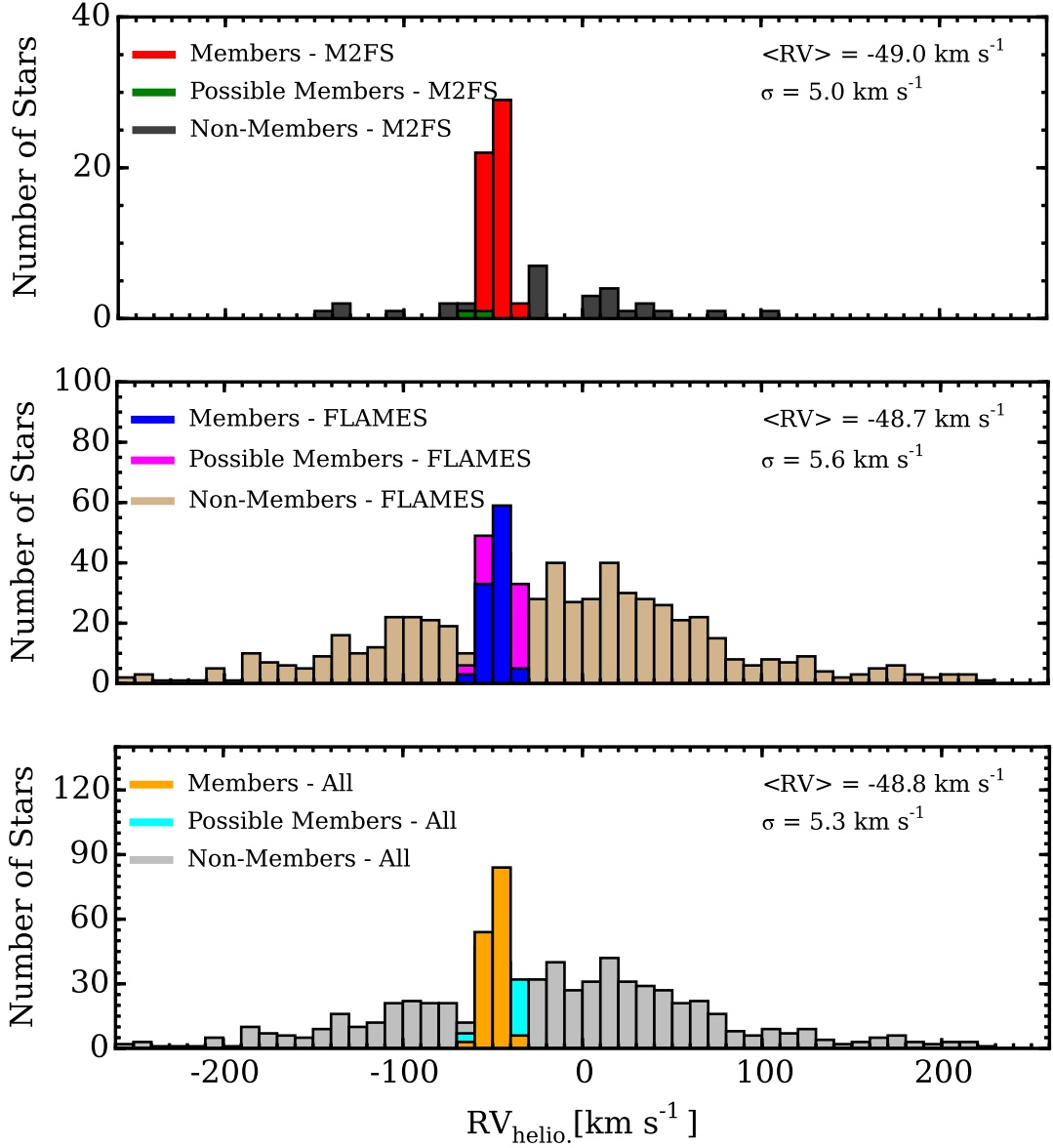}
\caption{The top, middle, and bottom panels illustrate the heliocentric radial
velocity distributions of member, possible member, and non--member stars 
observed with M2FS, FLAMES, and the combined sample, respectively.  For the 
bottom panel, we averaged the velocities for stars in common between the M2FS 
and FLAMES data sets.  All data are sampled into 10 km s$^{\rm -1}$ bins.
Note that the mean velocity and dispersion values only include member stars.}
\label{f3}
\end{figure}

Although \citet{Valenti11} found NGC 6569 to have a velocity dispersion of 
$\sim$8 km s$^{\rm -1}$, Figure \ref{f3} shows that cluster membership 
cannot be definitively determined based on radial velocity alone.  The 
contamination rate appears to be very small for the M2FS observations, but 
may be substantial for the FLAMES data set.  The cluster's mean velocity is 
clear in Figure \ref{f3} near --50 km s$^{\rm -1}$, but the local Galactic 
bulge velocity dispersion is at least 50--100 km s$^{\rm -1}$ 
\citep[e.g.,][]{Kunder12,Ness13,Zoccali14} and overlaps significantly with
the cluster distribution.  Therefore, the FLAMES target stars were
only identified as cluster members if their heliocentric radial velocities
were between --63 and --30 km s$^{\rm -1}$ ($\sim$3$\sigma$) \emph{and} their
[Fe/H] values were within $\sim$0.3 dex of the cluster's mean [Fe/H] $\sim$ 
--0.85 dex (see Section 5.1)\footnote{Most bulge field stars along NGC 6569's 
line--of--sight have [Fe/H] $\ga$ --0.8 dex \citep{Zoccali08}.}.  Stars that 
fell in the correct velocity range but did not have [Fe/H] determinations were 
labeled as possible members, and those with velocity and/or [Fe/H] measurements
that were inconsistent with cluster membership were labeled non--members.

For the M2FS RGB stars, we used the same membership criteria as for the FLAMES 
sample.  However, since we were unable to measure [Fe/H] for the HB stars, and
also for a few RGB stars, only the velocity values were used to assign 
membership in those cases.  The low contamination fraction shown in 
Figure \ref{f3} for the M2FS observations indicates that the false--positive 
membership rate should be low for the HB stars.  For the small number of stars
observed in both the M2FS and FLAMES setups, and for cases where [Fe/H] 
measurements were unavailable, we used the M2FS velocity data to assign
membership status.

Using the criteria outlined above and including only member stars, we found 
mean cluster velocities and dispersions of --49.0 km s$^{\rm -1}$ ($\sigma$ = 
5.0 km s$^{\rm -1}$), --48.7 km s$^{\rm -1}$ ($\sigma$ = 5.6 km s$^{\rm -1}$), 
and --48.8 km s$^{\rm -1}$ ($\sigma$ = 5.3 km s$^{\rm -1}$) for the M2FS, 
FLAMES, and combined data sets, respectively.  These values are in good 
agreement with \citet{Valenti11}, which measured a mean velocity of --47 $\pm$
4 km s$^{\rm -1}$.  Cluster members were found as far out as 9.9$\arcmin$ from 
NGC 6569's core, but 96$\%$ of the combined M2FS and FLAMES members were 
located inside the 6.9$\arcmin$ tidal radius \citep{Ortolani01}.  For the M2FS 
sample, the percentages of member, possible member, and non--member stars 
relative to the total sample were 64$\%$, 2$\%$, and 34$\%$, respectively.  
However, only 14$\%$ and 16$\%$ of the FLAMES sample included member and 
possible member stars while the remaining 70$\%$ were non--members.

\section{DATA ANALYSIS}

\subsection{Model Atmosphere Parameters: M2FS RGB Stars}

The model atmosphere parameters effective temperature (T$_{\rm eff}$), surface 
gravity (log(g)), metallicity ([Fe/H])\footnote{We used $\alpha$--enhanced
model atmospheres in order to account for differences between [M/H] and 
[Fe/H].}, and microturbulence ($\xi$$_{\rm mic.}$) were determined via 
spectroscopic methods for all RGB cluster members in the M2FS sample.  Since 
the M2FS HB and FLAMES HR13 data had lower S/N, resolution, and/or covered a 
much smaller wavelength region than the M2FS RGB data, we did not include a 
comprehensive composition analysis for those stars.  For the M2FS RGB sample,
T$_{\rm eff}$ values were derived by removing trends in plots of 
log $\epsilon$(\ion{Fe}{1}) versus lower excitation potential, and log(g) was 
estimated by enforcing ionization equilibrium between 
log $\epsilon$(\ion{Fe}{1}) and log $\epsilon$(\ion{Fe}{2}).  The 
microturbulence value for each star was set by removing trends in plots of 
log $\epsilon$(\ion{Fe}{1}) versus line strength, and model atmosphere 
metallicities were set to the mean of [\ion{Fe}{1}/H] and [\ion{Fe}{2}/H].
On average, the log $\epsilon$(\ion{Fe}{1}) and log $\epsilon$(\ion{Fe}{2})
abundances were based on measurements of 40 and 5 lines, respectively.

For all M2FS RGB stars, we assumed initial T$_{\rm eff}$, log(g), [Fe/H], and 
$\xi$$_{\rm mic.}$ values of 4200 K, 1.40 cgs, --0.80 dex, and
2.0 km s$^{\rm -1}$ and modified all four parameters simultaneously until a 
solution was found.  Since the ATLAS9 \citep{Castelli04} model atmosphere 
database is only available in increments of 250 K, 0.5 cgs, and 0.5 dex for 
T$_{\rm eff}$, log(g), and [Fe/H], we interpolated within the available 
grid\footnote{The ATLAS9 model atmosphere grids can be accessed at: 
http://wwwuser.oats.inaf.it/castelli/grids.html.} in order to derive the values
given in Table 4.  In principle, near--infrared colors from 2MASS and VVV could
be used to provide additional T$_{\rm eff}$ constraints.  However, we did not
employ this method for NGC 6569 because the J--K$_{\rm S}$ values for the RGB
stars analyzed here (see Table 2) are $\sim$0.1--0.2 magnitudes redder than 
the calibration range of most color--temperature relations 
\citep[e.g.,][]{Alonso99,Gonzalez09}.  Additionally, the line--of--sight 
reddening is moderately large for NGC 6569 with E(B--V) $>$ 0.55 magnitudes 
\citep{Zinn80,Bica83,Dutra00,Ortolani01,Valenti05}, and the VVV differential 
reddening map \citep{Gonzalez12} resolution is too coarse (2$\arcmin$) to 
provide dereddened magnitudes on a star--by--star basis.

\begin{figure}
\epsscale{0.75}
\plotone{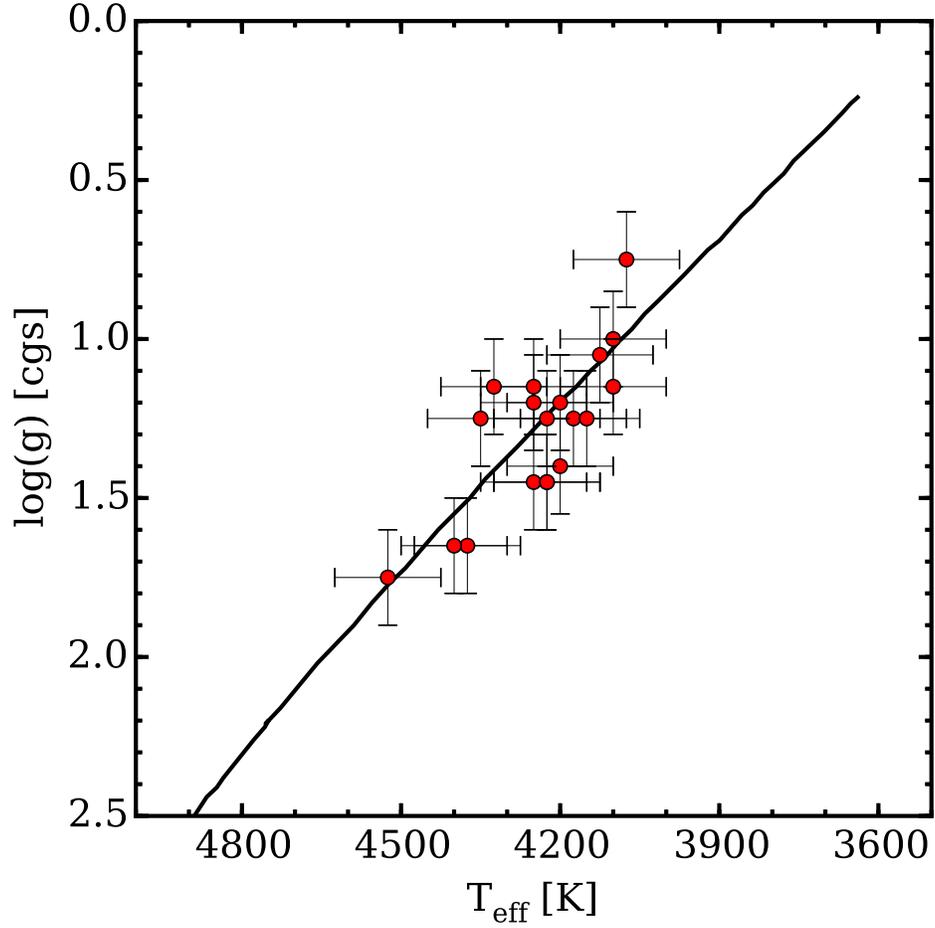}
\caption{The spectroscopically determined T$_{\rm eff}$ and log(g) values for
all M2FS stars are compared against a 10.9 Gyr $\alpha$--enhanced Dartmouth
isochrone \citep{Dotter08} with [Fe/H] = --0.85.  The error bars on each 
star represent the adopted T$_{\rm eff}$ and log(g) uncertainties adopted for
this analysis.  See text for details.}
\label{f4}
\end{figure}

Figure \ref{f4} compares our spectroscopic T$_{\rm eff}$ and log(g) values
against those predicted by a Dartmouth $\alpha$--enhanced isochrone 
\citep{Dotter08} assuming [Fe/H] = --0.85 (see Section 5.1) and an age of 
10.9 Gyr \citep{Santos04}, and shows that our adopted model atmosphere 
parameters are in good agreement with the isochrone predictions.  Specifically,
for a given measured T$_{\rm eff}$ value our log(g) determinations agree with 
the isochrone to within 0.02 cgs on average with a dispersion of 0.15 cgs.  
Similarly, for a given measured log(g) value our T$_{\rm eff}$ estimates agree 
with the isochrone to within 3 K on average with a dispersion of 80 K.  Note 
that a small number of stars with log(g) noticeably lower than the isochrone 
may belong to the AGB sequence.

\subsection{Equivalent Width and Spectrum Synthesis Analyses: M2FS RGB Stars}

Abundances of \ion{Fe}{1}, \ion{Fe}{2}, \ion{Si}{1}, \ion{Ca}{1}, \ion{Cr}{1},
and \ion{Ni}{1} were determined using the line list from \citet[][ see their 
Table 2]{Johnson15a}, the model stellar atmosphere parameters given in Table 4,
equivalent widths measured with a Gaussian profile deblending routine, and the 
\emph{abfind} driver of the local thermodynamic equilibrium (LTE) line analysis
code MOOG \citep{Sneden73}.  Although the abundances of Fe and other elements
may be susceptible to errors introduced by departures from LTE, 1D versus 3D
effects, and/or plane parallel versus spherical model atmosphere structure 
variations \citep[e.g.,][]{Lind11,Bergemann12,Dobrovolskas15}, we did not 
include any explicit corrections for these issues.  Instead, all abundances 
were measured relative to the cool, metal--poor giant Arcturus under the 
assumption that a differential analysis will cancel out most model atmosphere
deficiencies.  A list of the adopted Arcturus and solar log $\epsilon$(X)
abundances is provided in Table 2 of \citet{Johnson15a}.  The final 
[\ion{Fe}{1}/H], [\ion{Fe}{2}/H], [\ion{Si}{1}/Fe], [\ion{Ca}{1}/Fe],
[\ion{Cr}{1}/Fe], and [\ion{Ni}{1}/Fe] abundances for the M2FS RGB sample are
provided in Tables 5--6.

The abundances of \ion{O}{1}, \ion{Na}{1}, \ion{Mg}{1}, \ion{Al}{1}, 
\ion{La}{2}, and \ion{Eu}{2} were determined using the \emph{synth} driver in
MOOG.  Spectrum synthesis was required for these elements because their 
line profiles were affected by: nearby absorption lines, molecular equilibrium
calculations, autoionization lines, isotopic broadening, and/or hyperfine
broadening.  Given the interplay between the C, N, and O abundances in cool
stars and the ubiquity of CN lines in the 6140--6720 \AA\ region analyzed here,
we measured log $\epsilon$(O) first for all stars.  In particular, we set 
[C/Fe] = --0.3 dex and \iso{12}{C}/\iso{13}{C} = 5 following \citet{Valenti11},
and iteratively adjusted the log $\epsilon$(O) and log $\epsilon$(N) abundances
until a satisfactory fit was obtained for both the 6300.3 \AA\ [\ion{O}{1}] 
line and nearby CN features.  For oxygen and all other elements measured 
via spectrum synthesis, nearby atomic line log(gf) values were set via an
inverse Arcturus analysis assuming the abundances given in \citet{Ramirez11};
however, the CN line lists were adopted from \citet{Sneden14}.

Following a satisfactory determination of the oxygen abundance, and by 
extension the C $+$ N abundance, the 6154/6160 \AA\ and 6696/6698 \AA\ 
\ion{Na}{1} and \ion{Al}{1} doublets were fit for each star.  In cases where
we could not measure log $\epsilon$(O), CN line strengths were approximated 
using molecular line features near the Na and Al lines.  Mg abundances were
derived from the 6319 \AA\ triplet, and contributions from the overlapping
Ca autoionization line were estimated by examining the amount of continuum 
suppression present between $\sim$6316--6320 \AA.  The Ca autoionization line
strength was modified by altering the log $\epsilon$(Ca) abundance.  

Since many La lines in the 6140--6720 \AA\ window analyzed here exhibit signs
of significant hyperfine broadening, we fit the 6262 and 6390 \AA\ \ion{La}{2}
lines using the hyperfine structure line lists from \citet{Lawler01a}.  
Similarly, we fit the 6437 and 6645 \AA\ \ion{Eu}{2} lines using the line
lists from \citet{Lawler01b}, and assumed a solar isotope mixture of 47.8$\%$ 
and 52.2$\%$ for \iso{151}{Eu} and \iso{153}{Eu}, respectively.  However, we
did not include additional broadening effects due to isotope variations for 
La because each star's La abundance is expected to be dominated by 
\iso{139}{La}.  The adopted atomic parameters and reference Arcturus and solar 
abundances can be found in \citet{Johnson15a}, and a summary of the final
[X/Fe] ratios for all elements measured via spectrum synthesis is provided in
Tables 5--6.

\subsection{Calcium Triplet [Fe/H] Determinations}

Although we did not measure detailed chemical abundances for the FLAMES HR13
data, we did derive [Fe/H] values from the 8542 and 8662 \AA\ Calcium Triplet
lines of the HR21 spectra.  As noted by numerous previous authors
\citep[e.g.,][]{Armandroff101,Olszewski91,Idiart97,Rutledge97,Battaglia08},
the near--infrared CaT lines are reliable tracers of a star's metallicity
and are relatively insensitive to age and [Ca/Fe] abundance variations
\citep[e.g.,][]{Cole04,Carrera07,DaCosta16b}.  Following the methods outlined
in \citet{Yong16} and \citet{Johnson17a}, we employed the calibration 
described in \citet{Mauro14} to convert the measured CaT EWs into [Fe/H] 
abundances.  A key advantage of the \citet{Mauro14} CaT--[Fe/H] calibration
is that a star's luminosity parameter is defined as the difference between its 
K$_{\rm S}$--band magnitude and that of the HB.  For NGC 6569, which has 
E(B--V) $>$ 0.55 (see Section 4.1), a near--infrared calibration is preferred
over those using V--band magnitudes \citep[e.g.,][]{Starkenburg10,Saviane12,
Carrera13} because the K$_{\rm S}$--band is less sensitive to reddening.  As 
mentioned in Section 4.1, we do not have a high resolution differential
reddening map for NGC 6569 and have assumed a uniform reddening distribution.

In order to obtain [Fe/H] abundances from the CaT data, we first fit the 
8542 and 8662 \AA\ features using a function that represents the sum of a 
Gaussian and Lorentzian profile.  A summed EW ($\Sigma$EW) parameter is then
defined as:

\begin{equation}
\Sigma EW = EW_{8542} + EW_{8662},
\end{equation}

\noindent
and combined with Equation (3) in \citet{Mauro14} to give:

\begin{equation}
W^{\prime} = \Sigma EW - 0.385[K_{S}(HB) - K_{S}],
\end{equation}

\noindent
where W$^{\prime}$ represents the reduced EW and K$_{\rm S}$(HB) is the mean
magnitude of the red HB.  Since \citet{Mauro12} found evidence supporting red
HBs at K$_{\rm S}$ = 14.26 and 14.35, we have adopted K$_{\rm S}$ = 14.30 as
the mean cluster HB magnitude.  The final [Fe/H] abundances were determined 
using Equation (2) here and Equation (4) from \citet{Mauro14} to give:

\begin{equation}
[Fe/H] = -4.61 + 1.842\langle {W}'\rangle - 0.4428\langle {W}'\rangle^{2} + 0.04517\langle {W}'\rangle^{3}.
\end{equation}

\noindent
For cases in which both VVV and 2MASS K$_{\rm S}$ magnitudes were available, 
the VVV photometry was preferred.  Stars lacking both VVV and 2MASS photometry
were omitted from the CaT [Fe/H] analysis.  Table 7 provides a summary of the
individual CaT EWs, summed EWs, reduced EWs, and [Fe/H] determinations for all
cluster member stars.

\subsection{Abundance Uncertainties}

\subsubsection{M2FS RGB Stars}

Given the moderately high resolution and S/N of our data, the internal 
abundance uncertainties are dominated by errors in model atmosphere parameter
determinations.  For T$_{\rm eff}$, we have adopted an uncertainty value of 
100 K based on the log $\epsilon$(\ion{Fe}{1}) versus lower excitation 
potential plots, previous comparisons of spectroscopic and photometric 
temperatures using low reddening clusters \citep[e.g.,][]{Johnson15b,
Johnson17b}, and the scatter present in a plot of J--K$_{\rm S}$ versus 
spectroscopic temperature for the current data set.  Similarly, we have adopted
a conservative log(g) uncertainty of 0.15 cgs for all stars based on an 
examination of the scatter present when binning the data into groups spanning 
$\sim$100 K each (see also Section 4.1 and Figure \ref{f4}).  The mean 
line--to--line dispersion in [Fe/H] is 0.12 dex for \ion{Fe}{1} and 0.09 dex 
for \ion{Fe}{2}, and we have adopted a typical uncertainty of 0.10 dex for the 
model [M/H] value.  Finally, the $\xi$$_{\rm mic.}$ uncertainty was set at 0.10
km s$^{\rm -1}$ based on both an examination of the scatter present in plots of
log $\epsilon$(\ion{Fe}{1}) versus log(EW/$\lambda$) and the star--to--star 
dispersion in $\xi$$_{\rm mic.}$ for stars binned into 100 K groups.

The uncertainties in log $\epsilon$(X) were determined by varying each model
atmosphere parameter independently and measuring the difference in abundance
with the ``best--fit" value.  For species other than \ion{Fe}{1} and 
\ion{Fe}{2}, the [X/Fe] ratio uncertainties listed in Tables 5--6 take into
account the correlated variations in \ion{Fe}{1} and \ion{Fe}{2}.  In cases
where more than one line was used, a measurement uncertainty parameter 
defined as the standard deviation in log $\epsilon$(X) divided by the square 
root of the number of lines was included in the final uncertainty estimate.
For cases where only one line could be measured, a standard measurement 
uncertainty of 0.05 dex was included.  The abundance uncertainties listed in
Tables 5--6 represent the error sources listed above added in quadrature for 
each star.

\subsubsection{HR21 CaT Data}

For the HR21 data, the greatest source of uncertainty in the [Fe/H] 
determinations is the individual CaT EW measurements.  Following 
\citep{Johnson17a}, we estimated the profile fitting uncertainty by taking 
advantage of the strong correlation in EW between the 8542 and 8662 \AA\ lines,
and used the EW of each line to predict the expected EW of the other line.  The
new $\Sigma$EW values were then propagated through Equations 2--3 in Section 
4.3 to produce the predicted [Fe/H] abundances.  The mean difference in [Fe/H] 
between the predicted abundances and the value given in Table 7 was then 
adopted as the CaT metallicity uncertainty for each star.  We found a mean 
uncertainty of 0.15 dex ($\sigma$ = 0.11 dex), which is comparable to the 
fitting uncertainty of the \citep{Mauro14} calibration function.

Since one HR21 configuration was observed twice on nights separated by $\sim$1
month, we were able to perform an independent check of the CaT [Fe/H] 
uncertainty.  For the EWs, we found that the $\Sigma$EW values agreed between 
the two nights to within 4$\%$.  Additionally, we found a mean difference in 
[Fe/H] between the two observation sets to be 0.15 dex ($\sigma$ = 0.20 dex), 
which is compatible with our theoretical estimate.

Two remaining sources of uncertainty we did not account for are the cluster's
HB K$_{\rm S}$ magnitude and the K$_{\rm S}$ measurement errors for individual
stars.  For most stars, the VVV and/or 2MASS data have K$_{\rm S}$ measurement 
errors $\la$ 0.1 magnitudes and thus do not significantly affect the calibrated
metallicities.  In a similar sense, the $\sim$0.05 magnitude offset
between our adopted reference K$_{\rm S}$(HB) value and those of the HB--A and 
HB--B populations are unlikely to affect the [Fe/H] determinations at more
than the 0.03 dex level.  Differential reddening can also affect the 
K$_{\rm S}$(HB) -- K$_{\rm S}$ parameter in the metallicity calibration, but 
unless the reddening varies by more than a few tenths of a magnitude we can
safely ignore this effect.

\section{RESULTS AND DISCUSSION}

\subsection{Metallicity Distribution Function}

As mention in Section 1, the discovery of a bimodal HB in Terzan 5 has been
linked to a trimodal metallicity distribution that spans at least a factor of
10 in [Fe/H].  Although the K$_{\rm S}$--band HB spread in NGC 6569 noted by
\citet{Mauro12} is 0.2 magnitudes smaller than in Terzan 5, a possible cause
of the cluster's double HB is that NGC 6569 may also host stars spanning a 
wide range in [Fe/H].  However, we note that \citet{Munoz17} examined the 
composition of NGC 6440, which also showed evidence of a bimodal red HB, and
did not find evidence of a metallicity spread.

\begin{figure}
\epsscale{1.00}
\plotone{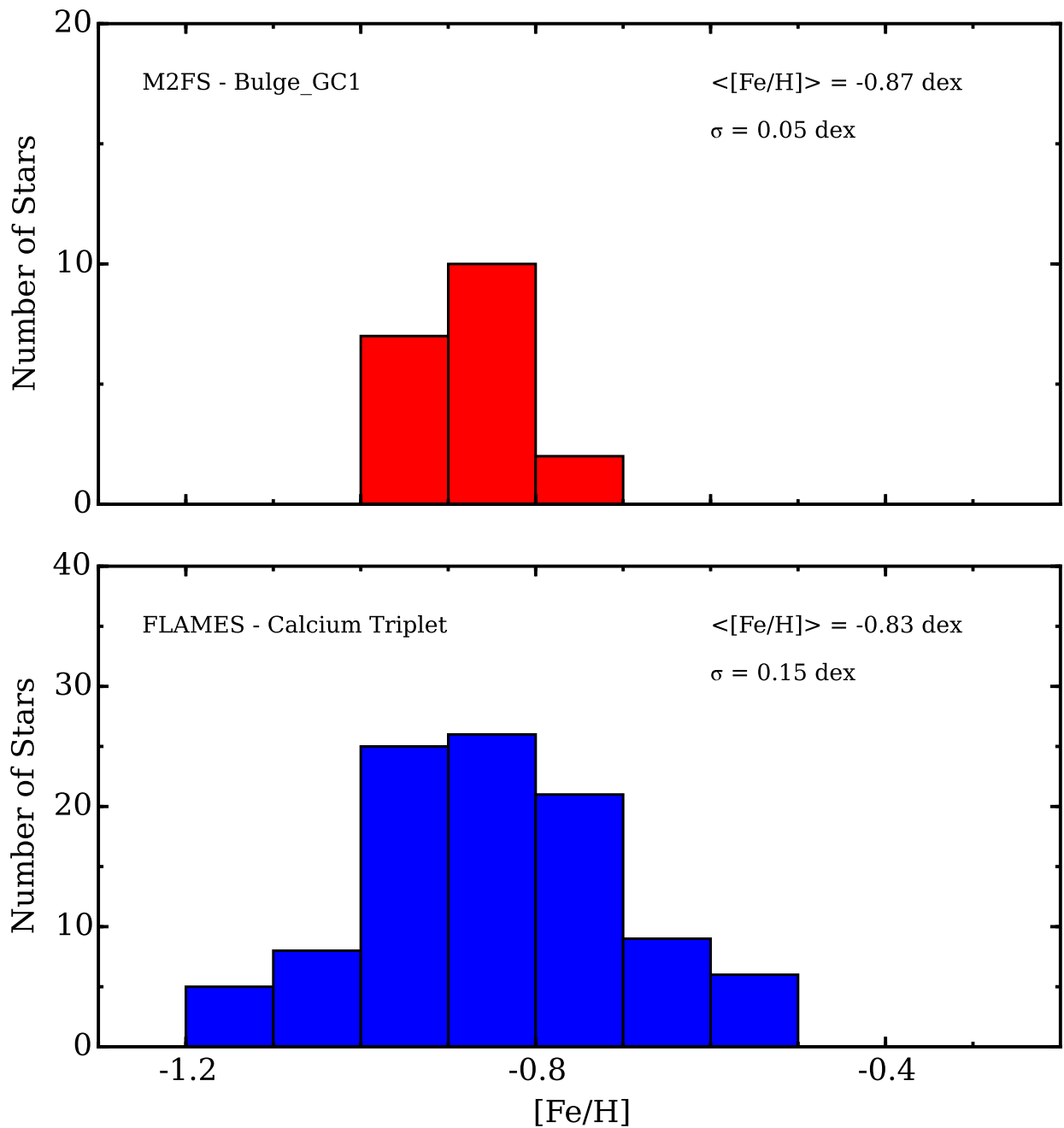}
\caption{The top and bottom panels illustrate the [Fe/H] distribution functions
for the M2FS and FLAMES samples, respectively.  The data are sampled into 
0.1 dex bins.  The mean uncertainty in [Fe/H] for the Calcium Triplet data
is 0.15 dex, which is comparable to the [Fe/H] dispersion shown here.  
Therefore, both the M2FS and FLAMES samples are consistent with NGC 6569 being 
a monometallic cluster.}
\label{f5}
\end{figure}

Figure \ref{f5} shows the results of our [Fe/H] measurements using both the 
M2FS RGB data and the FLAMES CaT observations.  Although the [Fe/H] dispersion 
is 0.15 dex for the CaT data, this value is equivalent to the mean measurement 
uncertainty determined in Section 4.4.2 for individual stars and is consistent 
with no intrinsic metallicity spread.  Similarly, the [Fe/H] dispersion 
determined from the M2FS RGB data is only 0.05 dex and is consistent with 
the [Fe/H] scatter found in other monometallic clusters 
\citep[e.g.,][]{Carretta09a}.  Neither data set provides evidence supporting 
the existence of a significant metallicity spread, and we conclude that 
NGC 6569 is a monometallic cluster.

For the M2FS RGB data and FLAMES CaT observations we find mean [Fe/H] 
abundances of --0.87 dex and --0.83 dex, respectively.  These values are 
consistent with previous spectroscopic and photometric estimates that ranged
from [Fe/H] $\sim$ --0.75 to --0.90 dex \citep{Zinn80,Bica83,Ortolani01,
Valenti05,Valenti11,Dias16}, and indicate that NGC 6569 is a relatively 
metal--rich globular cluster.

\subsection{Light Element Abundances}

\citet{Valenti11} represents the only high resolution spectroscopic analysis
of RGB stars in NGC 6569 and found moderate enhancements in [O/Fe] and [Al/Fe].
However, the small sample size (6 stars) of \citet{Valenti11} prevented a 
more detailed analysis, and they were not able to confirm whether the cluster
exhibits the usual light element abundance correlations.  The general
composition trends of the present analysis, based on 19 stars for most 
elements, are summarized in the box plot of Figure \ref{f6}, and indicate in
agreement with \citet{Valenti11} that at least [O/Fe] and [Al/Fe] are 
moderately enhanced with $\langle$[O/Fe]$\rangle$ = $+$0.44 dex ($\sigma$ = 
0.29 dex) and $\langle$[Al/Fe]$\rangle$ = $+$0.52 dex ($\sigma$ = 0.14 dex).
We also find that most stars are Na--enhanced with $\langle$[Na/Fe]$\rangle$ =
$+$0.13 dex ($\sigma$ = 0.21 dex), and that O, Na, and Al exhibit the largest
abundance ranges (0.6--0.8 dex) of all elements analyzed here.  

Figure \ref{f7} indicates that NGC 6569 exhibits the classical light element
abundance variations involving O, Na, and Al, such as the O--Na 
anti--correlation and Na--Al correlation, that are ubiquitous among old 
globular clusters \citep[e.g.,][]{Carretta09b,Carretta09c}.  However,
Figure \ref{f7} also shows that NGC 6569 does not exhibit any correlation
between [Mg/Fe] and [Al/Fe].  Additionally, the full abundance range in [Mg/Fe]
is more than a factor of two smaller than for [O/Fe] and the light odd--Z
elements, and with $\langle$[Mg/Fe]$\rangle$ = $+$0.41 dex ($\sigma$ = 0.09 
dex) Mg follows a pattern reminiscent of heavier $\alpha$--elements, such as
Ca (see Section 5.3).

Assuming the light element abundance variations in NGC 6569 are a result of 
``pristine" material mixing with gas that experienced high temperature 
proton--capture burning in a previous generation of more massive stars, 
the presence of clear O--Na and Na--Al (anti--)correlations coincident without 
a Mg--Al anti--correlation indicates that the burning temperatures were likely
in the range of $\sim$45--75 MK.  Temperatures lower than $\sim$45 MK would 
not have been able to produce Na and Al while those exceeding $\sim$75--100 MK
would have lead to significant \iso{24}{Mg} depletion
\citep[e.g.,][]{Prantzos07,DAntona16}. 

In general, clusters with higher masses and more extended blue HBs tend to 
exhibit signatures associated with more extreme light element processing 
(e.g., He enrichment; large O and Mg depletions; some Si production) and
higher burning temperatures \citep[e.g.,][]{Carretta07b,Yoon08,Milone14}.  
However, less advanced nuclear processing is expected as a cluster's 
metallicity increases due to effects such as: a general decline in the 
temperatures required to maintain hydrostatic equilibrium, enhanced mass loss, 
and an overall reduction in the range of light element yields from polluting 
stars \citep[e.g.,][]{Ventura09,Ventura13}.  Therefore, NGC 6569 follows a 
common trend among Galactic globular clusters with [Fe/H] $\ga$ --1 where a 
strong Mg--Al anti--correlation is only observed in the most massive clusters
that also contain significant populations of blue HB stars, such as NGC 6388 
and NGC 6441 \citep[e.g.,][]{Carretta09c}.

Figures \ref{f7} also shows evidence that NGC 6569 may host at least 
two stellar populations with distinct light element compositions.  Although 
we were only able to measure [O/Fe] for 8 stars, the O--Na panel of 
Figure \ref{f7} suggests a possible gap in the distribution near [O/Fe] $\sim$
$+$0.5 dex and [Na/Fe] $\sim$ $+$0.1 dex.  Similarly, Figure \ref{f8} plots 
[O/Na] as a function of [Al/H], which was shown by \citet{Johnson17c} to assist
in identifying discrete populations in globular clusters, and the data further 
reveal the possible presence of at least two populations.  For example, the gap
in [O/Na] between stars with [Al/H] $\sim$ --0.45 dex and those with [Al/H] 
$\sim$ --0.30 dex is at least a factor of 5, which is significantly larger than
the measurement uncertainties.  However, additional observations are required 
to determine how many chemically distinct populations exist in NGC 6569, and 
also to rule out the presence of stars with [O/Na] ratios between about $+$0.1 
and $+$0.6 dex.

\subsection{$\alpha$, Fe--Peak, and Neutron--Capture Abundances}

\begin{figure}
\epsscale{0.75}
\plotone{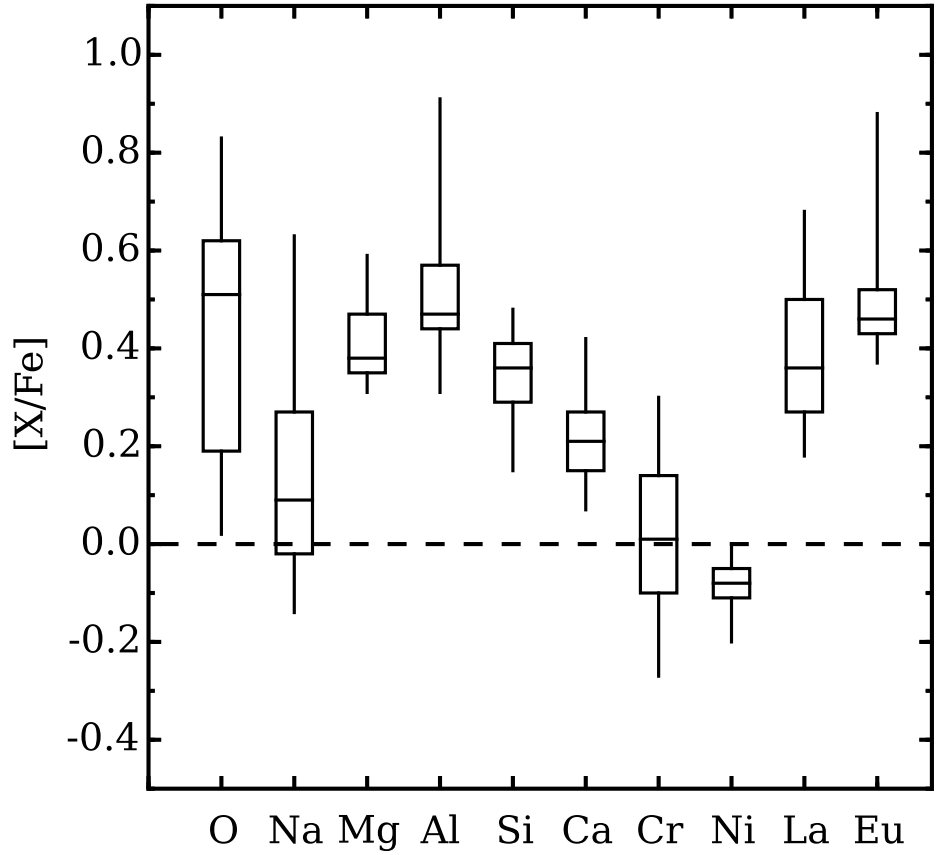}
\caption{A box plot is shown to illustrate the overall composition pattern of
NGC 6569.  For each element, the top, middle, and bottom horizontal lines 
represent the third quartile, median, and first quartile [X/Fe] abundances,
respectively.  Similarly, the vertical solid lines indicate the maximum and 
minimum [X/Fe] ratios of each element.}
\label{f6}
\end{figure}

\begin{figure}
\epsscale{0.80}
\plotone{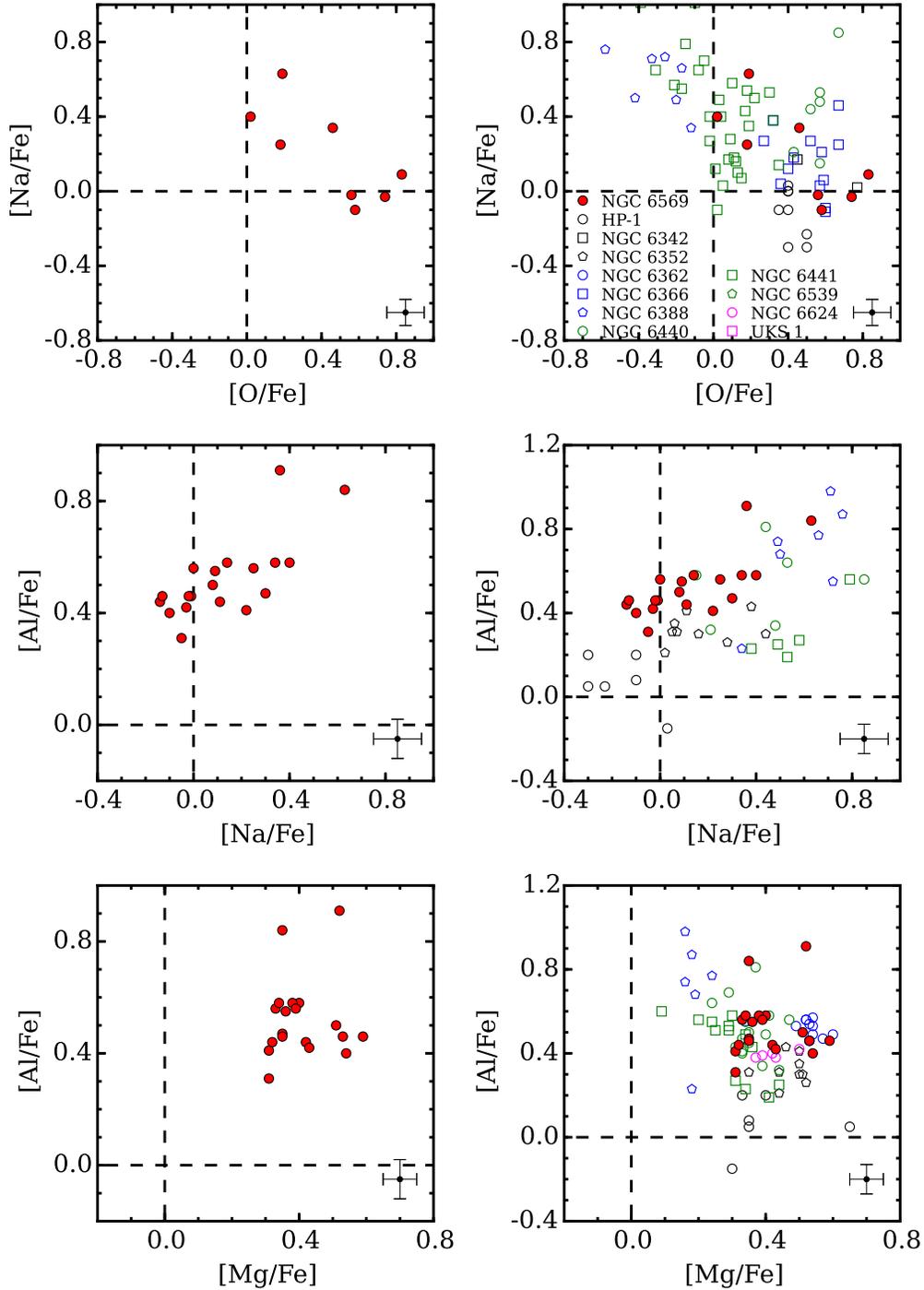}
\caption{Left: plots of [O/Fe] versus [Na/Fe], [Al/Fe] versus [Na/Fe], and 
[Al/Fe] versus [Mg/Fe] are shown for NGC 6569.  The data indicate an 
anti--correlation between [O/Fe] and [Na/Fe] and a correlation between [Na/Fe]
and [Al/Fe], but [Mg/Fe] and [Al/Fe] do not exhibit any correlation.  Right:
similar plots are shown comparing the light element [X/Fe] ratios of NGC 6569
against those of similar metallicity (--1.0 $\la$ [Fe/H] $\la$ --0.5) inner 
Galaxy ($\mid$l$\mid$ $\la$ 20$\degr$; $\mid$b$\mid$ $\la$ 20$\degr$; 
R$_{\rm GC}$ $\la$ 3 kpc) clusters.  The data sources are provided in Table 8.  
Representative error bars are included for each panel.}
\label{f7}
\end{figure}

\begin{figure}
\epsscale{0.75}
\plotone{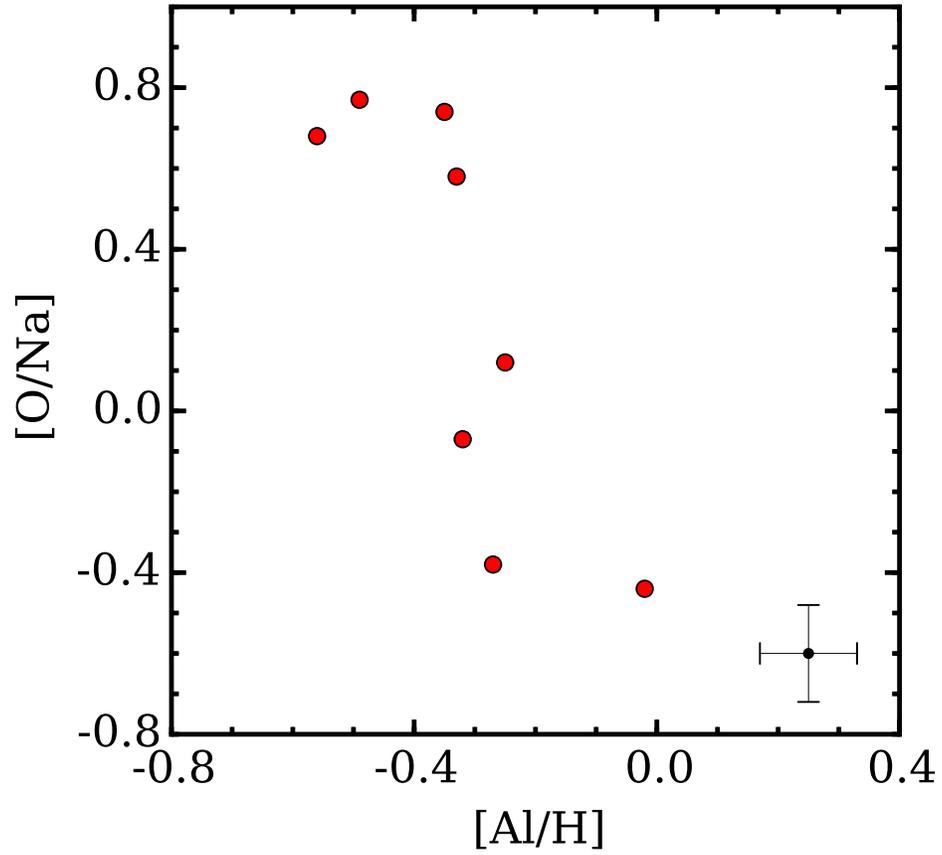}
\caption{[O/Na] is plotted as a function of [Al/H] for NGC 6569.  The data are
consistent with NGC 6569 hosting at least two distinct populations.}
\label{f8}
\end{figure}

As is shown in Figure \ref{f6}, we find that the heavier $\alpha$--elements
are enhanced in NGC 6569 with $\langle$[Si/Fe]$\rangle$ = $+$0.34 dex 
($\sigma$ = 0.09 dex) and $\langle$[Ca/Fe]$\rangle$ = $+$0.21 dex ($\sigma$ = 
0.10 dex).  The results presented here are in general agreement with those of
\citet{Valenti11}, but we derive mean [Si/Fe] and [Ca/Fe] ratios that are
lower by 0.15 dex and 0.10 dex, respectively.  However, the mean [Si/Ca] ratios
of the present work and \citet{Valenti11} agree to within 0.05 dex.  For the 
Fe--peak elements, we find approximately solar abundance ratios with 
$\langle$[Cr/Fe]$\rangle$ = $+$0.02 dex ($\sigma$ = 0.16 dex) and 
$\langle$[Cr/Fe]$\rangle$ = --0.08 dex ($\sigma$ = 0.05 dex).  Although the 
star--to--star dispersion in [Cr/Fe] is noticeably larger than for [Ni/Fe]
(see Figure \ref{f6}), we suspect that this is driven by larger measurement 
uncertainties rather than an astrophysical mechanism.

Figure \ref{f6} also shows that the neutron--capture elements are enhanced 
in NGC 6569 with $\langle$[La/Fe]$\rangle$ = $+$0.38 dex ($\sigma$ = 0.14 dex)
and $\langle$[Eu/Fe]$\rangle$ = $+$0.49 dex ($\sigma$ = 0.12 dex).  Similar 
to the case of Cr, the [La/Fe] and [Eu/Fe] dispersions are marginally larger
than those for elements such as Si, Ca, and Ni, but the interquartile ranges
(IQRs) of [La/Fe] and [Eu/Fe] are both smaller than for [Cr/Fe].  Additionally, 
an examination of the cluster's mean [La/Eu] composition, which is largely 
insensitive to surface gravity errors, indicates that NGC 6569 has a [La/Eu]
dispersion of 0.11 dex.  Since the star--to--star scatter in [La/Eu] is 
smaller than those of [La/Fe] and [Eu/Fe] individually, we conclude that the 
cluster's intrinsic heavy element spread does not exceed the $\sim$0.1 dex 
level.  NGC 6569 also has $\langle$[La/Eu]$\rangle$ = --0.11 dex, which 
suggests that the cluster's primordial composition was dominated by the 
r--process.  However, a mean [La/Eu] = --0.11 dex is a factor of 2--3 higher
than the lowest [La/Eu] values found in more metal--poor systems, and suggests
that the gas from which NGC 6569 formed may have also experienced some 
s--process enrichment (but see also Section 5.5 for an alternative 
interpretation).

\subsection{Comparing NGC 6569 with Other Galactic Bulge Globular Clusters}

The right panels of Figure \ref{f7} compare the [O/Fe], [Na/Fe], [Mg/Fe], and
[Al/Fe] abundances of individual stars in NGC 6569 against those in several
bulge/inner Galaxy clusters\footnote{In this context, bulge/inner Galaxy 
clusters are those with $\mid$l$\mid$ $\la$ 20$\degr$, $\mid$b$\mid$ $\la$ 
20$\degr$, and R$_{\rm GC}$ $\la$ 3 kpc.} with similar metallicities.  The data
indicate that O--Na anti-correlations and Na--Al correlations are common among 
bulge clusters with [Fe/H] $\ga$ --1, with only a few exceptions such as HP--1 
\citep{Barbuy06,Barbuy16} and NGC 6440 \citep{Munoz17} possibly lacking O--Na 
anti--correlations.  Figure \ref{f7} also shows that except for the peculiar 
clusters NGC 6388 and NGC 6441, which contain extended blue HBs, none of the
remaining bulge clusters within $\sim$0.3 dex of NGC 6569 exhibit Mg--Al
anti--correlations.  Therefore, NGC 6569 exhibits O--Na and Na--Al relations
that are more extended than most metal--rich bulge clusters, but does not reach
the most extreme levels of Na/Al--enhancement and O/Mg--depletion observed in 
NGC 6388 and NGC 6441.

Figures \ref{f9}--\ref{f10} compare the mean [X/Fe] ratios of NGC 6569 against
those of several bulge/inner Galaxy globular clusters spanning a wide [Fe/H]
range.  Although the mean [O/Fe], [Na/Fe], and [Al/Fe] abundances of the bulge 
cluster population exhibit considerable scatter, NGC 6569 follows the bulk 
trend by having enhanced mean [X/Fe] ratios of all three elements.  
Figures \ref{f9}--\ref{f10} also show that NGC 6569 follows the well--defined 
$\alpha$--element, Fe--peak, and neutron--capture element trends established 
by other bulge clusters.  From a bulk chemical perspective, the mean 
composition properties of NGC 6569 are indistinguishable from those of other
bulge clusters with similar metallicities.

\subsection{Comparing the Composition Patterns of Galactic Bulge Globular 
Clusters and Field Stars}

\begin{figure}
\epsscale{1.0}
\plotone{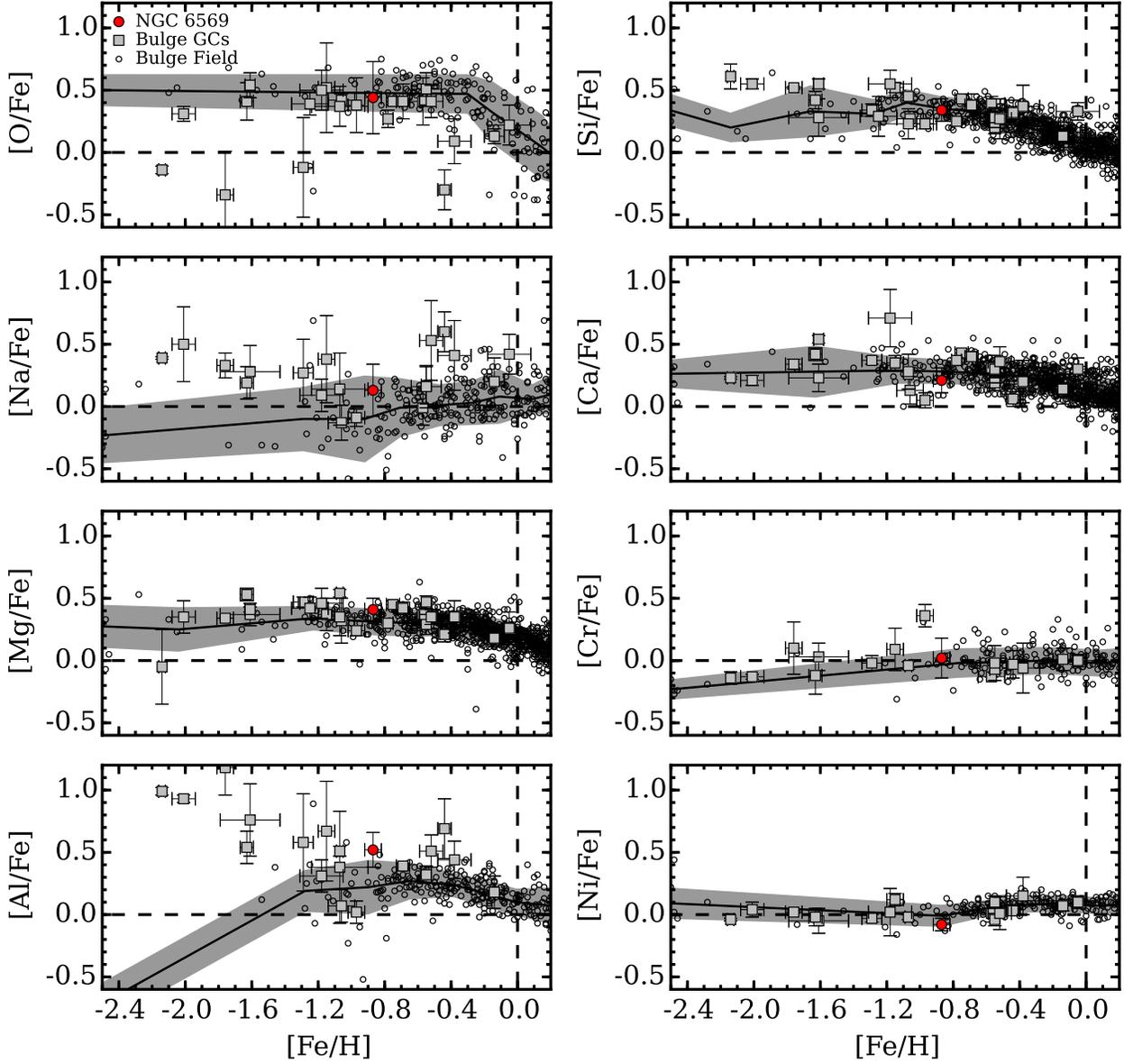}
\caption{The [X/Fe] ratios of elements ranging from O to Eu are plotted 
as a function of [Fe/H] for NGC 6569 (filled red circle), several Galactic 
bulge globular clusters (filled grey boxes), and a representative sample of
bulge field stars (open black circles).  The cluster symbols represent the 
mean [X/Fe] abundances and the error bars trace the standard deviation within
each cluster.  The solid black lines trace moving averages of the bulge 
field star data, and the filled dark grey regions illustrate the 1$\sigma$
variations.  The data sources are listed in Table 8.}
\label{f9}
\end{figure}

\begin{figure}
\epsscale{0.75}
\plotone{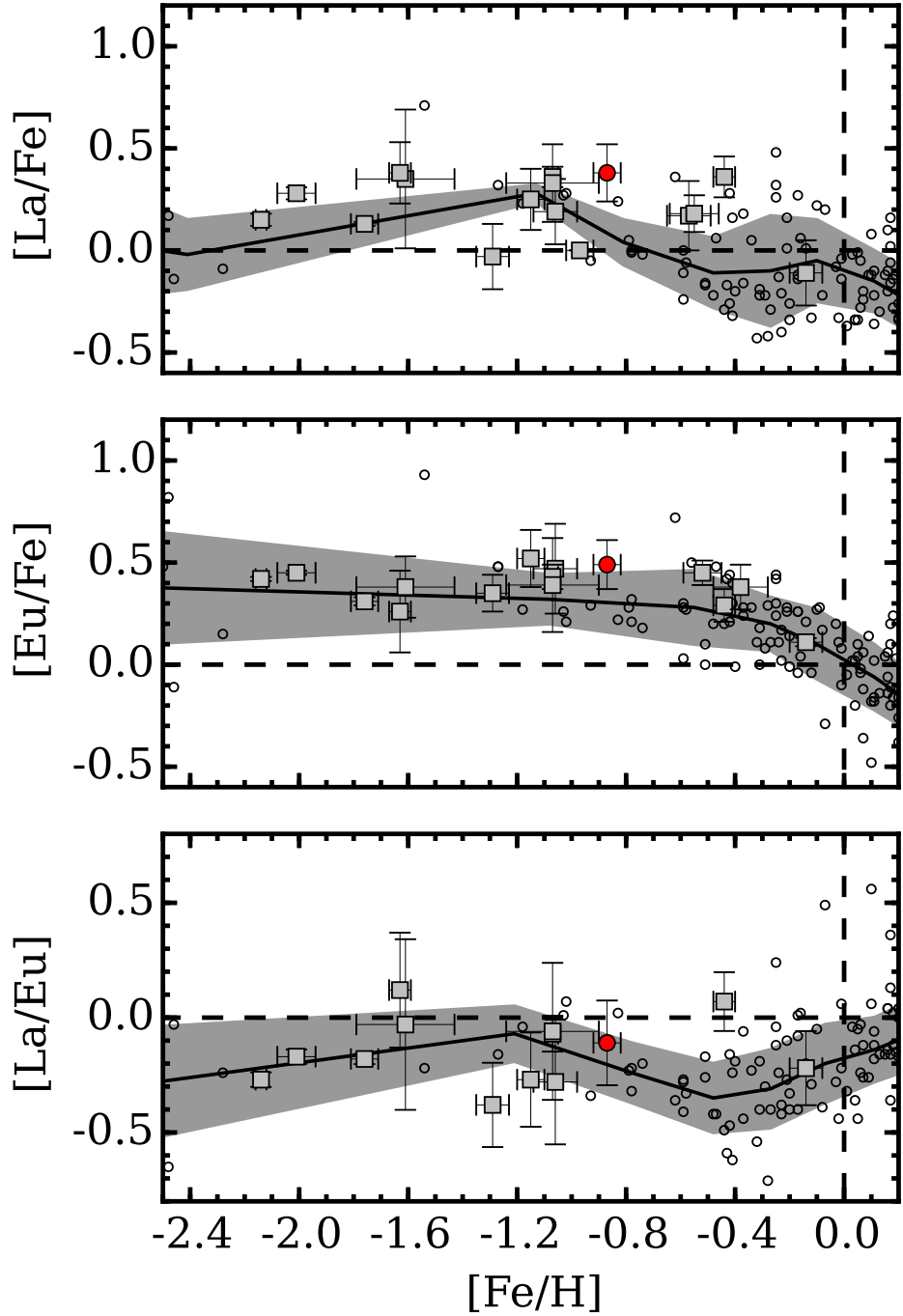}
\caption{Similar to Figure \ref{f9}, the [La/Fe], [Eu/Fe], and [La/Eu] ratios
of Galactic bulge clusters and field stars are plotted as a function of 
[Fe/H].}
\label{f10}
\end{figure}

Several dedicated studies have established the overall chemical composition
patterns of light, $\alpha$, Fe--peak, and heavy elements in the Galactic 
bulge \citep[e.g.,][]{McWilliam94,Fulbright07,Melendez08,AlvesBrito10,Ryde10,
Gonzalez11,Hill11,Rich12,Bensby13,Johnson14,VanDerSwaelmen16,Jonsson17}.
As is summarized in Figures \ref{f9}--\ref{f10}, bulge field stars generally
have: [Fe/H] $\ga$ --1 dex, enhanced [$\alpha$/Fe] and [Eu/Fe] for [Fe/H] $\la$
--0.4 dex, [X/Fe] $\sim$ 0 dex for Fe--peak elements, enhanced [La/Fe] that 
declines with increasing metallicity, and [La/Eu] ratios between about --0.5 
and --0.1 dex with a possible increase at super--solar metallicities.  
Furthermore, the bulge field stars exhibit [Na/Fe] abundances that either
slowly increase with [Fe/H] or present a ``zig--zag" pattern associated with
competition between core--collapse and thermonuclear supernovae 
\citep{McWilliam16}, and [Al/Fe] traces the general trends of [$\alpha$/Fe] and
[Eu/Fe].  These observations indicate that a significant fraction of the bulge
formed rapidly, experienced chemical enrichment dominated by massive 
stars, and likely had a star formation rate that was a few times higher than
the local thick disk.  However, the lowest metallicity ([Fe/H] $\la$ --1.5 dex)
bulge field stars exhibit chemical composition patterns that may be more 
similar to those found in metal--poor thick disk and halo stars 
\citep{GarciaPerez13,Howes14,Howes15}.

Figures \ref{f9}--\ref{f10} compare the mean composition patterns of several
bulge/inner Galaxy globular clusters against those of the bulge field stars.  
With the exception of [O/Fe], [Na/Fe], and [Al/Fe], Figures \ref{f9}--\ref{f10} 
indicate that the mean [$\alpha$/Fe], Fe--peak, and heavy element compositions 
of the bulge globular cluster and field stars are similar, and perhaps 
indistinguishable for many elements, over a wide range in metallicity.  A 
particularly interesting question is whether the bulge clusters remain 
$\alpha$--enhanced to a higher metallicity than the field stars, but 
unfortunately the number of globular clusters studied at [Fe/H] $\ga$ --0.3 dex
is too small to draw any strong conclusions.  For example, NGC 6528 is the most
metal--rich cluster shown and exhibits mean [Mg/Fe] and possibly [Ca/Fe] 
abundances that are similar to the field, but the cluster may have a mean 
[Si/Fe] ratio that is $\sim$0.2 dex higher.  However, the clusters that are 
only slightly more metal--poor than NGC 6528 have [$\alpha$/Fe] ratios that 
are within the range observed for similar metallicity bulge field stars.

In a similar sense, Figure \ref{f10} shows some ambiguity regarding the [La/Fe]
ratios for bulge field and cluster stars with [Fe/H] $\ga$ --0.8 dex.  Both
\citet{Johnson12} and \citet{VanDerSwaelmen16} noted a general decrease in
[La/Fe] with increasing [Fe/H], and also found stars with [La/Fe] $\sim$ $+$0.2
dex at [Fe/H] $\ga$ --0.6 dex; however, \citet{Johnson12} found most stars 
with [Fe/H] between --0.8 and 0.0 dex to have [La/Fe] $<$ 0 dex while 
\citet{VanDerSwaelmen16} found most stars in that metallicity range to have
[La/Fe] $>$ 0 dex.  As a result, the data do not clearly differentiate between
whether or not the bulge clusters near [Fe/H] $\sim$ --0.6 dex have elevated
mean [La/Fe] ratios compared to the field stars or are within the normal 
range.  The bulge clusters exhibit a similar decrease in [La/Fe] with 
increasing [Fe/H] observed in the field stars, but the metallicity at which
the downturn occurs lies between the results of \citet{Johnson12} and 
\citet{VanDerSwaelmen16}.  

Interestingly, the [Eu/Fe] trends in Figure \ref{f10} are nearly identical 
for the bulge cluster and field stars, and the two populations may share 
similar [La/Eu] distributions as well.  The available [La/Eu] data indicate 
significant contributions by the r--process for both the cluster and field 
star compositions, but both populations have mean [La/Eu] ratios that are
higher than the --0.6 dex pure r--process limit observed in some metal--poor 
halo and globular cluster stars \citep[e.g.,][]{Roederer10}.  The enhanced 
[La/Eu] ratios in bulge cluster and field stars are likely a result of global
s--process enrichment in the Galaxy driven by pollution from AGB stars 
\citep[e.g.,][]{Gallino98,Bisterzo10} and/or massive ``spinstars" 
\citep[e.g.,][]{Frischknecht16}.  In particular, previous data have shown that 
most Galactic cluster and field stars with [Fe/H] $\ga$ --2 dex exhibit at 
least some evidence of s--process enrichment, such as [La/Eu] ratios that 
increase with metallicity \citep[e.g.,][]{James04,Simmerer04,DOrazi10}.  

Alternatively, we note \citet{Roederer10} found that pure r--process 
halo/disk cluster and field stars with [Fe/H] $<$ --1.4 can have --0.6 $\la$ 
[La/Eu] $\la$ --0.05 dex, which suggests the moderately depleted [La/Eu] ratios
in the bulge could still represent nearly pure r--process compositions, at least
for [Fe/H] $<$ 0 dex.  This may be especially true for the bulge clusters shown
in Figure \ref{f10}, which have $\langle$[La/Eu]$\rangle$ = --0.14 dex ($\sigma$
= 0.15 dex) and fail to exhibit any variations with metallicity.  However, a 
predominantly r--process origin would require a high star formation rate in 
order to mitigate s--process pollution from $\la$ 4 M$_{\rm \sun}$ AGB stars.

While the bulge cluster and field stars may share similar mean [La/Eu] trends, 
Figure \ref{f9} shows that significant differences are found when examining
the [O/Fe], [Na/Fe], and [Al/Fe] abundance trends.  For example, the bulge 
clusters exhibit larger cluster--to--cluster variations in mean light element
composition than are observed among field stars of similar [Fe/H].  Several 
clusters also contain large numbers of stars with lower [O/Fe] and higher 
[Na,Al/Fe] abundances than are found in the field, and most clusters exhibit
clear O--Na anti--correlations and Na--Al correlations that reflect strong
self--enrichment.  Notably, the light element (anti--)correlations prevalent
in globular clusters are not found in bulge field stars\footnote{We note that
[Na/Fe] and [Al/Fe] are correlated for intermediate metallicity bulge stars, 
but the [Na/Fe] and [Al/Fe] ranges are smaller than observed in globular 
clusters and are not accompanied by similar O--Na anti--correlations.  The 
bulge Na--Al correlation is likely driven by the similar production mechanisms 
of Na and Al in massive stars, and does not reflect the same proton--capture
nucleosynthesis processes that operate in cluster environments.}, which 
indicates that a large fraction of the bulge cannot have originated from 
dissolved globular clusters hosting the same chemical properties and population
ratios as those in Figure \ref{f9}.  We note that a population of N/Al--rich 
stars has been found recently in the inner Galaxy \citep{Schiavon17b}, but the 
metallicity distribution of these stars peaks near [Fe/H] $\sim$ --1 dex.  As a
result, their progenitor cluster systems are too metal--poor to have built--up
the bulge field star population.  However, self--enriched but now dissolved
clusters could have contributed a small percentage of stars to the bulge's 
total mass.  

\subsection{Insight into NGC 6569's Double Horizontal Branch}

As mentioned in Section 1, NGC 6569 is particularly interesting because 
\citet{Mauro12} found evidence that the cluster, along with NGC 6440, may host 
two red HBs separated in the K$_{\rm S}$--band by $\sim$0.1 mag.  An intriguing
possibility raised by \citet{Mauro12} is that split red HBs are common in 
massive, metal--rich bulge clusters, and that similar to Terzan 5 all bulge
clusters with double HBs may host two or more stellar populations with 
different metallicities.  However, RGB composition analyses have so far failed
to detect intrinsic metallicity spreads in both NGC 6440 \citep{Origlia08,
Munoz17} and NGC 6569 \citep[][this paper]{Valenti11}.

In order to gain additional insight regarding the origin of the double HB in 
NGC 6569, we have derived radial velocities and CaT metallicities of stars 
in each HB group defined by \citet{Mauro12}.  Additionally, we have combined
the bluer but lower S/N M2FS and FLAMES spectra of individual stars in each
HB population to create co--added spectra that will permit a search for mean
light element variations between the two HBs.  Although \citet{Mauro12} found
double red HBs in NGC 6440 and NGC 6569, similar observations of the 
metal--rich bulge clusters NGC 6380, NGC 6441, NGC 6528, and NGC 6553 did not 
reveal similarly complex red HBs.  Therefore, it is prudent to confirm that the 
HB--A and HB--B populations both contain cluster members.  

\begin{figure}
\epsscale{1.00}
\plotone{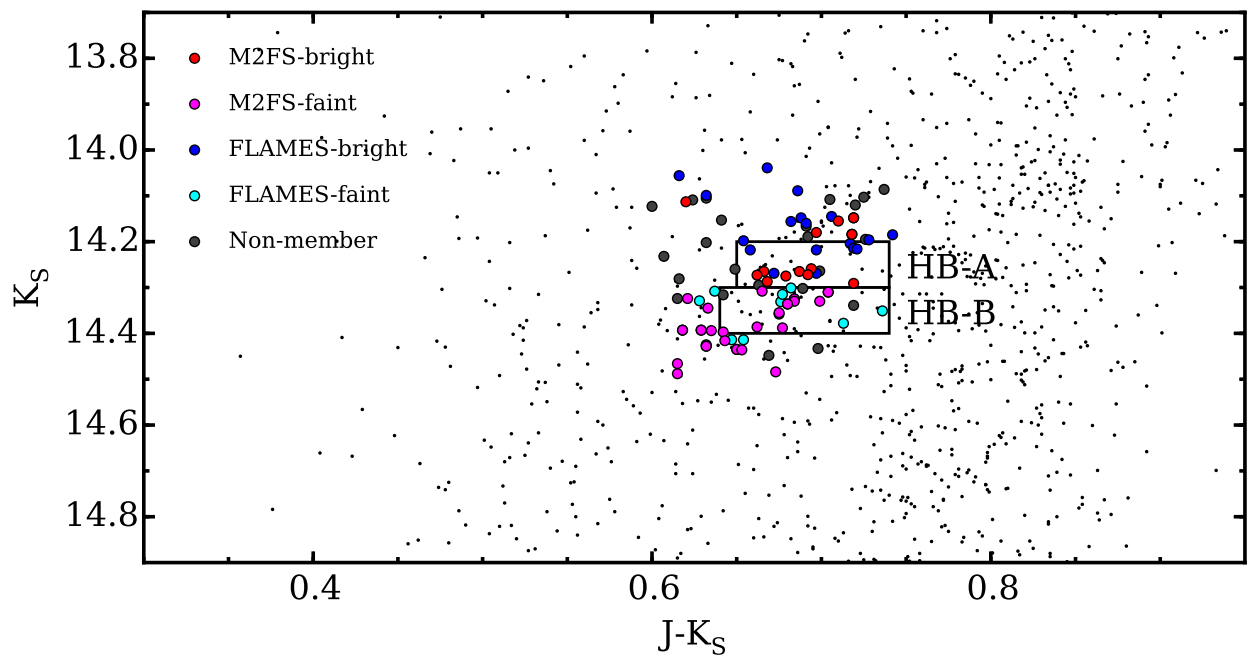}
\caption{A VVV K$_{\rm S}$ versus J--K$_{\rm S}$ color--magnitude diagram is
shown for the HB region of NGC 6569.  The two large boxes indicate the 
magnitude and color ranges where \citet{Mauro12} identified a possible double
HB.  The filled red and magenta circles indicate stars observed with M2FS that
may belong to the ``bright" (HB--A) and ``faint" (HB--B) populations, 
respectively.  Similarly, the filled blue and cyan circles show stars observed 
with FLAMES that may belong to the HB--A and HB--B populations.  The filled
grey circles indicate stars with radial velocities that are inconsistent with 
cluster membership.}
\label{f11}
\end{figure}

The HB targets analyzed in the present data set are shown in Figure \ref{f11}, 
and indicate that both the HB--A and HB--B groups identified by 
\citet{Mauro12} contain cluster members.  If we restrict the examination to
include only stars located within the selection boxes of Figure \ref{f11},
then the data indicate that the mean heliocentric radial velocities of the 
HB--A and HB--B groups are identical within the errors.  Specifically, the
M2FS HB--A and HB--B populations have mean velocities of --49.5 km s$^{\rm -1}$
$\pm$ 1.2 km s$^{\rm -1}$ and --48.8 km s$^{\rm -1}$ $\pm$ 1.5 km s$^{\rm -1}$, respectively, while the FLAMES HB--A and HB--B populations have mean velocities
of --46.7 km s$^{\rm -1}$ $\pm$ 4.5 km s$^{\rm -1}$ and --45.9 km s$^{\rm -1}$ 
$\pm$ 2.5 km s$^{\rm -1}$, respectively.  

\begin{figure}
\epsscale{0.75}
\plotone{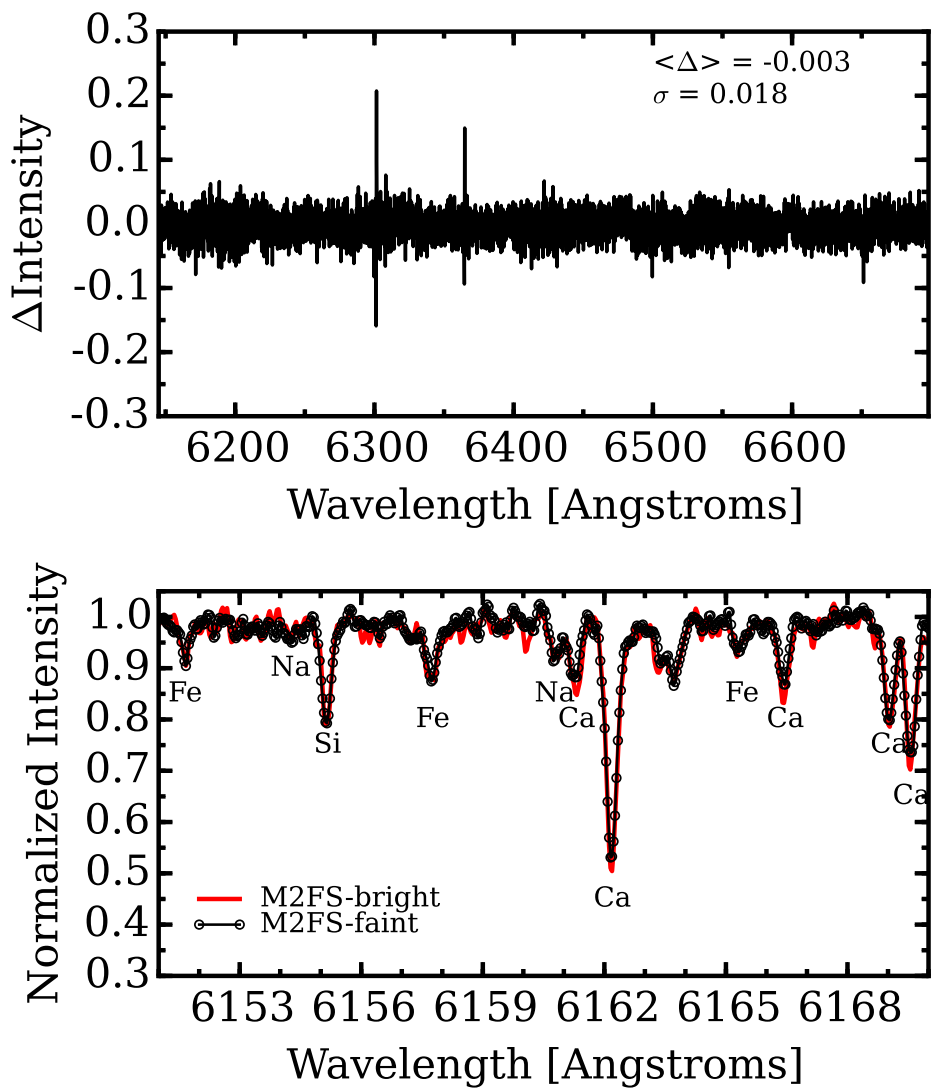}
\caption{Top: after co--adding all M2FS spectra for stars residing in the 
bright and faint HB boxes of Figure \ref{f11}, this panel shows the difference 
in normalized line strength ($\Delta$Intensity) between the mean bright and 
faint HB populations as a function of wavelength.  Except for spectral regions 
near the 6300 and 6363 \AA\ telluric features, the two populations have mean 
spectra that agree to within 1.8$\%$.  Bottom: a sample of the spectral region
near the 6154/6160 \ion{Na}{1} lines compared the co--added M2FS--bright (red 
lines) and M2FS--faint (open black circles and lines) populations.  The data 
do not present strong evidence favoring [Fe/H], [$\alpha$/Fe], or light element 
abundance variations as the cause of the double HB.}
\label{f12}
\end{figure}

\begin{figure}
\epsscale{0.75}
\plotone{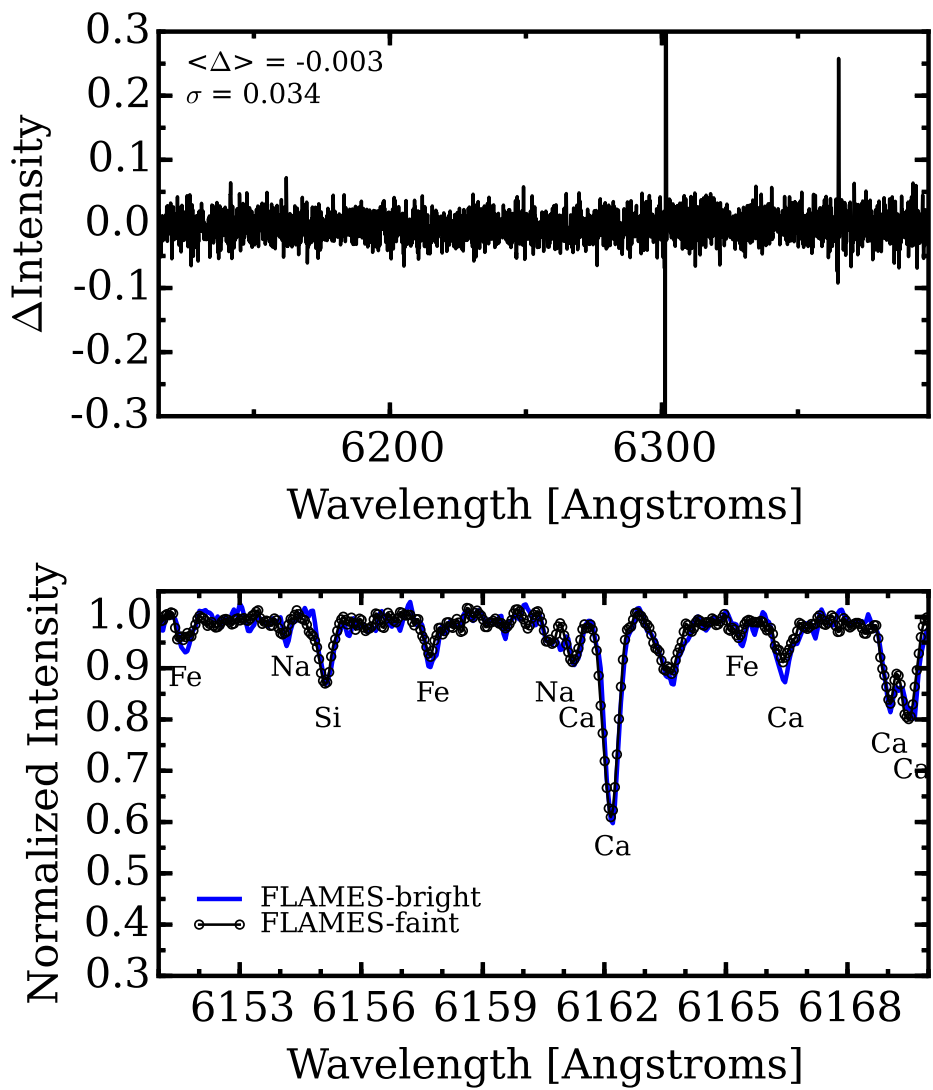}
\caption{Top: similar to Figure \ref{f12}, the difference in line strength 
between the co--added FLAMES HR13 spectra of bright and faint HB stars is 
plotted as a function of wavelength.  The mean spectra agree to within 3.4$\%$.
Bottom: a comparison of the spectral region near the 6154/6160 \ion{Na}{1} 
lines shows that the bright and faint HB stars exhibit similar mean [Fe/H],
[$\alpha$/Fe], and light element abundances.}
\label{f13}
\end{figure}

Having confirmed that the HB--A and HB--B groups contain cluster members, we 
can now investigate possible composition differences between the two 
populations.  Using the FLAMES CaT data, and again restricting the observations
to include only stars within the selection boxes of Figure \ref{f11}, we find 
the HB--A group to have a mean [Fe/H] = --0.76 dex ($\sigma$ = 0.07 dex) and 
the HB--B group to have a mean [Fe/H] = --0.89 dex ($\sigma$ = 0.05 dex).  
Although the brighter K$_{\rm S}$--band magnitude of the HB--A population 
is consistent with a higher mean metallicity and matches the pattern observed 
in Terzan 5, we suspect that the 0.13 dex difference in mean metallicity 
between the HB--A and HB--B groups is not significant.  For example, the mean 
CaT measurement uncertainty is $\sim$0.15 dex (see Section 4.4.2 and Table 7) 
for individual stars, and the HB metallicity estimates are based on only 
$\sim$5 stars in each group.  Additionally, the metallicity distribution 
functions shown in Figure \ref{f5} are not bimodal, and the RGB line--by--line 
abundance analysis produced an [Fe/H] dispersion of only 0.05 dex, despite 
spanning the full color range of the RGB (see Figure \ref{f1}).  A comparison 
of the co--added HB--A and HB--B spectra in Figures \ref{f12}--\ref{f13} also 
reveals that their line strengths agree to within 1.8--3.4$\%$, which leaves 
little room for significant composition differences.  Therefore, we conclude 
that the double HB in NGC 6569 is not driven by an intrinsic metallicity 
spread.

Assuming the age dispersion within NGC 6569 is negligible, He enhancement is 
an additional parameter that can produce a luminosity dispersion on the HB
\citep[e.g.,][]{Valcarce12}.  Although we cannot directly measure He abundances
with the present data set, extreme He variations can be traced in globular 
clusters by searching for large light element abundance variations 
\citep[e.g.,][]{Bragaglia10a,Bragaglia10b,Dupree11}.  \citet{Mauro12} noted 
that He enhancements of Y $>$ 0.30 could be required to explain the HB 
luminosity variations in at least NGC 6440, and if this is the cause for the
double HB feature in NGC 6569 too then we should be able to detect higher mean 
Na abundances in the He--enhanced HB stars.  

Figure \ref{f7} indicates that if the HB--A and HB--B populations exhibit 
different light element abundances then the expected lower limit in 
$\Delta$[Na/Fe] between the two groups should be $\sim$0.4 dex.  For a typical
red HB star with [Fe/H] $\sim$ --0.85 dex, a [Na/Fe] difference of 0.4 dex 
translates to a central line depth difference of $\sim$5--7$\%$ for the 
6154/6160 \AA\ \ion{Na}{1} lines in R $\sim$ 27,000 spectra.  However, 
comparisons of the co--added HB--A and HB--B M2FS and FLAMES spectra in 
Figures \ref{f12}--\ref{f13} show that, at least on average, the 
\ion{Na}{1} lines vary by $\la$ 2$\%$ in depth.  Therefore, the HB--A and HB--B
populations likely have mean [Na/Fe] abundances that agree to within about 
0.2--0.3 dex.  Although a detailed analysis of individual stars should be 
carried out for confirmation, with the present co--added spectra we conclude 
that the HB--A and HB--B populations do not possess significantly different 
mean [Na/Fe] abundances.  As a result, He mass fraction differences exceeding
$\Delta$Y = 0.05--0.10 are unlikely.

\begin{figure}
\epsscale{0.75}
\plotone{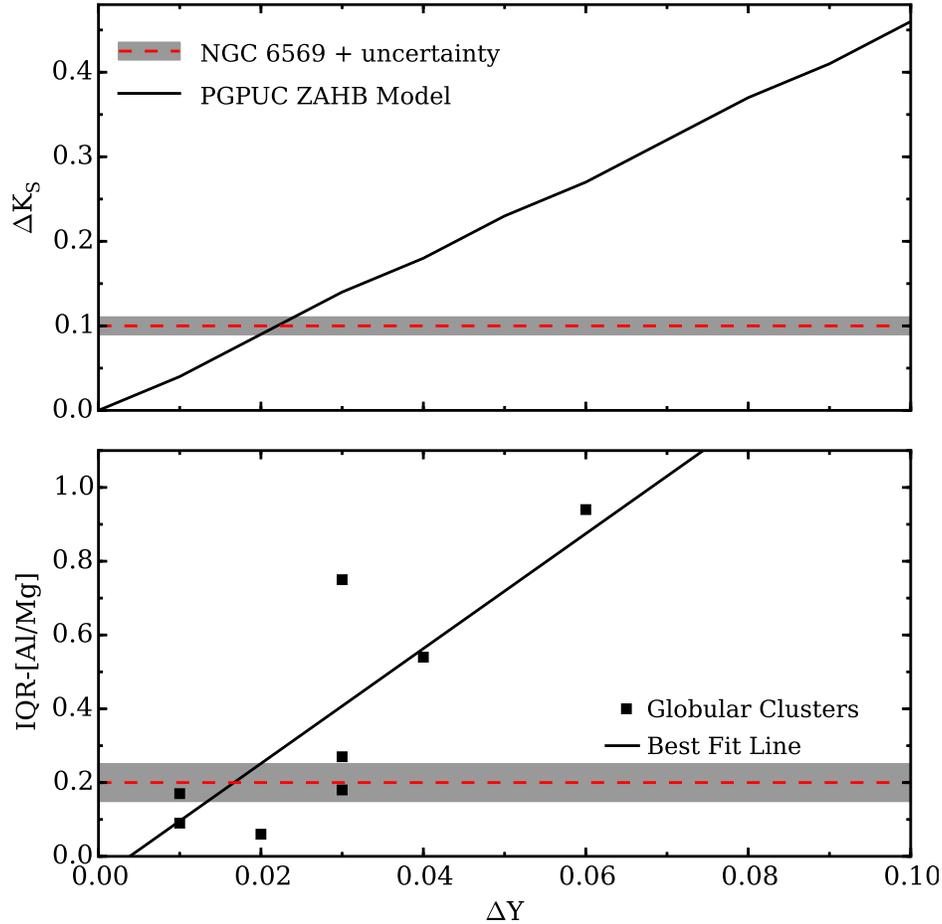}
\caption{\emph{Top:} the solid black line shows the predicted change in a 
0.7 M$_{\rm \odot}$ red HB star's K$_{\rm S}$--band magnitude 
($\Delta$K$_{\rm S}$) on the zero age HB (ZAHB) as a function of He enhancement
($\Delta$Y) where $\Delta$Y = 0 corresponds to Y = 0.25.  The isochrone model 
was obtained from the Princeton--Goddard--PUC (PGPUC) stellar evolution 
database \citep{Valcarce12} assuming an age of 10.9 Gyr, 
[Fe/H] = --0.87 dex, [$\alpha$/Fe] = $+$0.3 dex, and metallicity (Z) values 
that scale with changes in Y.  The dashed red line indicates the 
measured difference in K$_{\rm S}$ between the two red HBs in NGC 6569, and the
grey shaded region indicates the $\sim$0.01 magnitude uncertainty.  
\emph{Bottom:} the observed change in a cluster's [Al/Mg] interquartile range
(IQR) as a function of $\Delta$Y is shown using the compilation of 
\citet{Gratton10}.  The red dashed line indicates the [Al/Mg] IQR for NGC 6569,
and the shaded uncertainty region was calculated by drawing 10$^{\rm 3}$ 
random samples from the observed [Al/Mg] distribution and determining the 
standard deviations of the IQR values.  Note that the \citet{Gratton10} 
$\Delta$Y values refer to the difference between a cluster's median and maximum
He abundances.  The combined panels indicate that $\Delta$Y $\sim$ 0.02 may be 
sufficient to produce a ZAHB K$_{\rm S}$--band difference of $\sim$0.1 
magnitudes, and that the [Al/Mg] spread in NGC 6569 is compatible with 
$\Delta$Y $\sim$ 0.01--0.03, at least when compared to other clusters in the
\citet{Gratton10} compilation.}
\label{f14}
\end{figure}

We conclude by investigating the possibility that a small He abundance 
variation could both be present in NGC 6569 and be responsible for the double 
red HB feature.  The top panel of Figure \ref{f14} illustrates the expected 
difference in K$_{\rm S}$ between two zero age HB (ZAHB) stars with similar 
ages, metallicities, and [$\alpha$/Fe] abundances but with various differences 
in He mass fraction ($\Delta$Y)\footnote{The isochrone models used in 
Figure \ref{f14} are from the Princeton--Goddard--PUC (PGPUC) database 
\citep{Valcarce12}, which can be accessed at 
http://www2.astro.puc.cl/pgpuc/index.php.  A significant limitation of the 
present investigation is that the comparison stars are assumed to reach the 
RGB--tip with the same mass (0.7 M$_{\rm \odot}$; the PGPUC grid lower HB mass 
limit); however, He--rich stars are likely to evolve faster and have a lower 
ZAHB mass.  Additional effects such as mass loss and rotation are also not 
considered.}.  Under these assumptions, Figure \ref{f14} indicates that the 
HB--A and HB--B populations could exhibit 0.1 magnitude differences in 
K$_{\rm S}$ if the two groups differed in Y by $\sim$0.02--0.025.  Is such a He
difference compatible with the observations?  \citet{Gratton10} showed that a 
correlation exists between a cluster's [Al/Mg] IQR and $\Delta$Y, and the 
bottom panel of Figure \ref{f14} compares these data against similar IQR 
measurements for NGC 6569 RGB stars\footnote{An important caveat in this 
analysis is that the [Al/Mg] IQR versus $\Delta$Y correlation shown in 
Figure \ref{f14} is primarily based on clusters with [Fe/H] $<$ --1.  Higher
metallicity clusters tend to have smaller [Al/Fe] spreads 
\citep[e.g.,][see their Figure 3]{Carretta09c} so the $\Delta$Y value 
estimated here for NGC 6569 may be a lower limit.}.  Figure \ref{f14} shows 
that NGC 6569's [Al/Mg] IQR of 0.20 $\pm$ 0.05 dex is compatible with a 
$\Delta$Y value of $\sim$ 0.015--0.03.  Although the present data do not 
provide a clear explanation for the origin of the double red HB feature in 
NGC 6569, we can conclude that the light element abundance spreads in NGC 6569 
are consistent with other clusters for which $\Delta$Y reaches the 0.02 
threshold necessary to separate ZAHB stars by $\sim$ 0.1 magnitudes in the
K$_{\rm S}$--band.

\section{SUMMARY}

We utilized new and archival high resolution (R $\sim$ 27,000) spectra from the
Magellan--M2FS and VLT--FLAMES spectrographs to investigate the radial 
velocities, light and heavy element abundances, and CaT metallicities of
RGB and HB stars located near the Galactic bulge globular cluster NGC 6569.  
We derived a mean cluster heliocentric radial velocity of --48.8 km 
s$^{\rm -1}$ ($\sigma$ = 5.3 km s$^{\rm -1}$), but recommend using both 
velocity and metallicity measurements to establish
membership because the cluster's systemic velocity overlaps significantly with
the bulge field distribution.  Fortunately, the cluster has a mean [Fe/H] 
$\approx$ --0.85 dex, which is more metal--poor than most bulge field stars.
We note that the M2FS 6140--6720 \AA\ data and FLAMES CaT data yielded [Fe/H] 
dispersions of only 0.05 and 0.15 dex, respectively, which are consistent with
NGC 6569 being a monometallic cluster.

The light element abundance distributions of NGC 6569 follow the typical 
patterns observed in old globular clusters.  For example, the [O/Fe], [Na/Fe],
and [Al/Fe] abundances exhibit full ranges of $\sim$0.6--0.8 dex, and the 
cluster exhibits a clear O--Na anti--correlation and Na--Al correlation.  
However, [Mg/Fe] and [Si/Fe] exhibit dispersions of only 0.09 dex each, and 
neither abundance ratio is correlated with [O/Fe], [Na/Fe], or [Al/Fe].  The 
data are therefore consistent with a scenario in which second generation 
(O--poor; Na/Al--rich) stars formed from gas that was processed at temperatures
of $\sim$45--75 MK.  We also find some evidence that NGC 6569 may be decomposed
into at least two distinct populations with different light element 
compositions.

The $\alpha$, Fe--peak, and neutron--capture element abundances are generally 
consistent with rapid formation and chemical enrichment.  The [X/Fe] ratios 
of Mg, Si, and Ca are all enhanced by about a factor of two relative to the 
Sun, and the star--to--star scatter is $\la$ 0.1 dex for each element.  Both 
[Cr/Fe] and [Ni/Fe] exhibit mean abundances that are approximately solar with
dispersions of 0.16 and 0.05 dex, respectively.  
We suspect that the larger [Cr/Fe] dispersion is due to increased measurement 
uncertainties rather than intrinsic cosmic scatter.  For the neutron--capture 
elements, we find $\langle$[La/Fe]$\rangle$ = $+$0.38 dex ($\sigma$ = 0.14 dex)
and $\langle$[Eu/Fe]$\rangle$ = $+$0.49 dex ($\sigma$ = 0.12 dex), and the 
moderately depleted mean [La/Eu] ratio of --0.11 dex suggests significant 
pollution via the r--process.  However, NGC 6569 does not reach the lowest 
[La/Eu] values found in some halo clusters, and as a result may have 
experienced some s--process enrichment.

The overall light and heavy element chemical composition patterns of NGC 6569 
match the mean trends exhibited by other bulge/inner Galaxy globular clusters.
However, the O--Na anti--correlation and Na--Al correlation in NGC 6569 
extend to some of the lowest [O/Fe] and highest [Na,Al/Fe] values found in
metal--rich ([Fe/H] $\ga$ --1 dex) bulge clusters, but do not reach the most
extreme values observed in clusters such as NGC 6388 and NGC 6441.  As a 
population, the bulge clusters exhibit little change in their mean [La/Eu] 
abundances from at least [Fe/H] $\sim$ --2.2 to --0.15 dex, which suggests 
s--process enrichment has been minimal.  In fact, the mean heavy element 
composition of $\langle$[La/Eu]$\rangle$ = --0.14 dex ($\sigma$ = 0.15 dex) 
for the bulge clusters is within the pure r--process range observed in 
metal--poor ([Fe/H] $<$ --1.4 dex) halo stars and clusters.

A comparison of mean compositions between bulge/inner Galaxy globular cluster
and field stars revealed that both populations exhibit similar abundance 
trends for a wide range of elements.  For [Fe/H] $\la$ --0.4 dex, the 
[$\alpha$/Fe], [Cr/Fe], [Ni/Fe], [Eu/Fe], and [La/Eu] distributions are nearly
indistinguishable between the two groups.  At [Fe/H] $\ga$ --0.4 dex, more data
are needed to determine if the clusters remain $\alpha$--enhanced to higher
[Fe/H] than the field stars, and also to determine if the two populations 
share similar [La/Fe] trends.  Clear light element abundance differences are 
present between the bulge cluster and field stars for [O/Fe], [Na/Fe], and 
[Al/Fe] across a wide range in [Fe/H].  Specifically, the clusters scatter to 
higher [Na,Al/Fe] and lower [O/Fe], and the field stars do not exhibit the same
light element (anti--)correlations that are prevalent in the globular clusters.
Despite sharing similar $\alpha$, Fe--peak, and heavy element abundances, the 
light elements rule out that a large fraction of bulge field stars could have
originated from self--enriched but now dissolved globular clusters.

Following the discovery of a double red HB in NGC 6569 by \citet{Mauro12}, we 
investigated the radial velocity, metallicity, and light element abundances 
of each population.  The velocity and [Fe/H] measurements indicate that both 
HBs contain cluster members, but unlike the case of Terzan 5 we did not detect 
a large metallicity spread.  We found the brighter HB to have
a mean [Fe/H] that is higher by 0.13 dex, but we do not consider the [Fe/H]
difference to be significant given the small samples sizes ($\sim$5 stars in
each HB group), the $\sim$0.15 dex CaT measurement uncertainties of individual
stars, and the small 0.05 dex [Fe/H] dispersion observed for the M2FS RGB 
sample.  A further comparison of the co--added spectra revealed that the 
line strengths vary by $\la$ 3$\%$ between the two HB groups, and we did not
detect significant differences in mean light element composition.  By 
extension, we infer that both HB populations have similar He abundances.
However, we cannot rule out He abundance differences as small as $\Delta$Y
$\sim$ 0.02 that may be sufficient to reproduce the observed double red HB
feature.

\acknowledgements

This research has made use of NASA's Astrophysics Data System Bibliographic
Services.  This research has made use of the services of the ESO Science 
Archive Facility.  This publication has made use of data products from the Two 
Micron All Sky Survey, which is a joint project of the University of 
Massachusetts and the Infrared Processing and Analysis Center/California 
Institute of Technology, funded by the National Aeronautics and Space 
Administration and the National Science Foundation.  C.I.J. gratefully 
acknowledges support from the Clay Fellowship, administered by the Smithsonian 
Astrophysical Observatory.  M.M. is grateful for support from the National 
Science Foundation to develop M2FS (AST--0923160) and carry out the 
observations reported here (AST--1312997), and to the University of Michigan 
for its direct support of M2FS construction and operation.  M.G.W. is supported
by National Science Foundation grants AST--1313045 and AST--1412999.  R.M.R 
acknowledges support from grant AST--1413755 from the National Science 
Foundation.

\clearpage

\clearpage

\clearpage

\clearpage

\clearpage

\clearpage

\clearpage

\clearpage

\clearpage

\clearpage

\clearpage

\clearpage

\clearpage

\clearpage

\clearpage
\input{tab1.tex}
\clearpage
\input{tab2.tex}
\clearpage
\setlength{\voffset}{0.75in}
\input{tab3.tex}
\clearpage
\input{tab4.tex}
\clearpage
\setlength{\voffset}{0.75in}
\input{tab5.tex}
\clearpage
\setlength{\voffset}{0.75in}
\input{tab6.tex}
\clearpage
\setlength{\voffset}{0.75in}
\input{tab7.tex}
\clearpage
\input{tab8.tex}

\end{document}

%% file: tab1.tex
\tablenum{1}
\tablecolumns{5}
\tablewidth{0pt}

\begin{deluxetable}{ccccc}
\tablecaption{Observation Log}
\tablehead{
\colhead{Field Number}     &
\colhead{Instrument}      &
\colhead{Filter}      &
\colhead{UT Date}      &
\colhead{Exposure Time}	\\
\colhead{}	&
\colhead{}      &
\colhead{}      &
\colhead{}      &
\colhead{(s)}      
} 

\startdata
1	&	M2FS	&	Bulge$\_$GC1	&	2014 June 03	&	3 $\times$ 2400	\\
2	&	FLAMES	&	HR13	&	2014 July 03	&	1 $\times$ 2775	\\
2	&	FLAMES	&	HR13	&	2014 August 01	&	1 $\times$ 2775	\\
3	&	FLAMES	&	HR21	&	2015 June 22	&	1 $\times$ 1800	\\
4	&	FLAMES	&	HR21	&	2015 June 27	&	1 $\times$ 2700	\\
5	&	FLAMES	&	HR21	&	2015 June 27	&	1 $\times$ 2700	\\
6	&	FLAMES	&	HR21	&	2015 June 28	&	1 $\times$ 2700	\\
7	&	FLAMES	&	HR21	&	2015 July 26	&	1 $\times$ 1800	\\
8	&	FLAMES	&	HR21	&	2015 July 26	&	1 $\times$ 1800	\\
6	&	FLAMES	&	HR21	&	2015 July 26	&	1 $\times$ 2700	\\
9	&	M2FS	&	Bulge$\_$GC1	&	2016 June 28	&	3 $\times$ 1800	\\
9	&	M2FS	&	Bulge$\_$GC1	&	2016 June 30	&	3 $\times$ 1800	\\
9	&	M2FS	&	Bulge$\_$GC1	&	2016 July 01	&	3 $\times$ 2000	\\
9	&	M2FS	&	Bulge$\_$GC1	&	2016 July 01	&	1 $\times$ 1800	\\
\enddata

\tablecomments{Observations with matching field numbers indicate that the same
fiber and instrument configurations were used.}

\end{deluxetable}

%% file: tab2.tex
\tablenum{2}
\tablecolumns{12}
\tablewidth{0pt}

\begin{deluxetable}{cccccccccccc}
\rotate
\tabletypesize{\tiny}
\tablecaption{M2FS Star Identifiers, Coordinates, Photometry, and Velocities}
\tablehead{
\colhead{Star Name}	&
\colhead{RA}      &
\colhead{DEC}      &
\colhead{J}      &
\colhead{K$_{\rm S}$}      &
\colhead{Star Name}      &
\colhead{RA}      &
\colhead{DEC}      &
\colhead{J}      &
\colhead{K$_{\rm S}$}      &
\colhead{RV$_{\rm helio.}$}      &
\colhead{RV Error}      \\
\colhead{(2MASS)}      &
\colhead{(J2000)}      &
\colhead{(J2000)}      &
\colhead{(mag)}      &
\colhead{(mag)}      &
\colhead{(VVV)}      &
\colhead{(J2000)}      &
\colhead{(J2000)}      &
\colhead{(mag)}      &
\colhead{(mag)}      &
\colhead{(km s$^{\rm -1}$)}      &
\colhead{(km s$^{\rm -1}$)}
} 

\startdata
\multicolumn{12}{c}{Cluster Members}       \\
\hline
18125997$-$3154025	&	273.249915	&	$-$31.900700	&	15.021	&	14.285	&	J181259.99$-$315402.48	&	273.249967	&	$-$31.900692	&	14.952	&	14.265	&	$-$52.6	&	0.3	\\
18133083$-$3148103\tablenotemark{a}	&	273.378474	&	$-$31.802876	&	11.759	&	10.628	&	J181330.83$-$314810.39	&	273.378486	&	$-$31.802886	&	11.737	&	10.565	&	$-$48.7	&	0.2	\\
18133123$-$3149306	&	273.380141	&	$-$31.825172	&	15.076	&	14.308	&	J181331.23$-$314930.67	&	273.380129	&	$-$31.825189	&	14.955	&	14.287	&	$-$52.0	&	0.4	\\
18133181$-$3149104	&	273.382578	&	$-$31.819578	&	15.109	&	14.343	&	J181331.82$-$314910.62	&	273.382596	&	$-$31.819618	&	15.014	&	14.330	&	$-$57.6	&	0.4	\\
18133187$-$3148259	&	273.382802	&	$-$31.807209	&	15.010	&	14.361	&	J181331.87$-$314825.98	&	273.382799	&	$-$31.807219	&	14.931	&	14.265	&	$-$42.7	&	0.3	\\
18133187$-$3150175\tablenotemark{a}	&	273.382816	&	$-$31.838215	&	14.457	&	13.917	&	J181331.90$-$315017.56	&	273.382939	&	$-$31.838213	&	14.902	&	14.184	&	$-$60.6	&	0.4	\\
18133222$-$3147158\tablenotemark{a}	&	273.384260	&	$-$31.787737	&	12.526	&	11.434	&	J181332.22$-$314715.86	&	273.384264	&	$-$31.787739	&	12.497	&	11.426	&	$-$49.3	&	0.2	\\
\enddata

\tablenotetext{a}{Observed with both M2FS and FLAMES.}

\tablecomments{The first set of coordinates and photometry are on the 2MASS
system and the second set are on the VVV system.}

\end{deluxetable}

%% file: tab3.tex
\tablenum{3}
\tablecolumns{15}
\tablewidth{0pt}

\begin{deluxetable}{ccccccccccccccc}
\rotate
\tabletypesize{\tiny}
\tablecaption{FLAMES Star Identifiers, Coordinates, Photometry, and Velocities}
\tablehead{
\colhead{Star Name}	&
\colhead{RA}      &
\colhead{DEC}      &
\colhead{J}      &
\colhead{K$_{\rm S}$}      &
\colhead{Star Name}      &
\colhead{RA}      &
\colhead{DEC}      &
\colhead{J}      &
\colhead{K$_{\rm S}$}      &
\colhead{Star Name}	&
\colhead{RA}	&
\colhead{DEC}	&
\colhead{RV$_{\rm helio.}$}      &
\colhead{RV Error}      \\
\colhead{(2MASS)}      &
\colhead{(J2000)}      &
\colhead{(J2000)}      &
\colhead{(mag)}      &
\colhead{(mag)}      &
\colhead{(VVV)}      &
\colhead{(J2000)}      &
\colhead{(J2000)}      &
\colhead{(mag)}      &
\colhead{(mag)}      &
\colhead{(VLT)}      &
\colhead{(J2000)}      &
\colhead{(J2000)}      &
\colhead{(km s$^{\rm -1}$)}      &
\colhead{(km s$^{\rm -1}$)}
} 

\startdata
\multicolumn{15}{c}{Cluster Members}       \\
\hline
18131138$-$3148128	&	273.297457	&	$-$31.803564	&	11.104	&	9.921	&	J181311.39$-$314812.91	&	273.297464	&	$-$31.803588	&	11.034	&	9.753	&	3002910	&	273.297	&	$-$31.8036	&	$-$47.8	&	0.7	\\
18132157$-$3151036	&	273.339884	&	$-$31.851017	&	14.125	&	13.269	&	J181321.57$-$315103.68	&	273.339885	&	$-$31.851023	&	14.024	&	13.148	&	3001489	&	273.340	&	$-$31.8510	&	$-$40.4	&	0.6	\\
18132381$-$3151202	&	273.349243	&	$-$31.855625	&	12.237	&	11.177	&	J181323.81$-$315120.31	&	273.349248	&	$-$31.855644	&	12.185	&	11.155	&	3001323	&	273.349	&	$-$31.8556	&	$-$52.4	&	0.6	\\
18132416$-$3151095	&	273.350701	&	$-$31.852650	&	14.167	&	13.292	&	\nodata	&	\nodata	&	\nodata	&	\nodata	&	\nodata	&	3001259	&	273.351	&	$-$31.8526	&	$-$44.9	&	0.7	\\
18132703$-$3150191	&	273.362656	&	$-$31.838642	&	14.063	&	13.102	&	J181327.03$-$315019.17	&	273.362653	&	$-$31.838660	&	13.980	&	13.120	&	3000849	&	273.363	&	$-$31.8386	&	$-$52.5	&	0.9	\\
18132719$-$3155450	&	273.363333	&	$-$31.929184	&	10.640	&	9.786	&	J181327.19$-$315545.14	&	273.363331	&	$-$31.929206	&	10.588	&	9.867	&	3003469	&	273.363	&	$-$31.9292	&	$-$40.2	&	0.6	\\
18132987$-$3147575	&	273.374495	&	$-$31.799328	&	14.487	&	13.547	&	J181329.88$-$314757.53	&	273.374508	&	$-$31.799314	&	14.435	&	13.584	&	2000503	&	273.374	&	$-$31.7993	&	$-$43.1	&	0.8	\\
\enddata

\tablenotetext{a}{Observed with both M2FS and FLAMES.}

\tablecomments{The first set of coordinates and photometry are on the 2MASS
system, the second set are on the VVV system, and the third set of identifiers
and coordinates are from the FLAMES image headers.}

\end{deluxetable}

%% file: tab4.tex
\tablenum{4}
\tablecolumns{6}
\tablewidth{0pt}

\begin{deluxetable}{cccccc}
\tablecaption{M2FS RGB Cluster Member Model Atmosphere Parameters}
\tablehead{
\colhead{Star Name}     &
\colhead{Star Name}      &
\colhead{T$_{\rm eff}$}      &
\colhead{log(g)}      &
\colhead{[Fe/H]}	&
\colhead{$\xi$$_{\rm mic.}$}	\\
\colhead{(2MASS)}	&
\colhead{(VVV)}      &
\colhead{(K)}      &
\colhead{(cgs)}      &
\colhead{(dex)}      &
\colhead{(km s$^{\rm -1}$)}
} 

\startdata
18133083$-$3148103	&	J181330.83$-$314810.39	&	4075	&	0.75	&	$-$0.91	&	2.05	\\
18133222$-$3147158	&	J181332.22$-$314715.86	&	4125	&	1.05	&	$-$0.96	&	1.70	\\
18133324$-$3150194	&	J181333.24$-$315019.50	&	4225	&	1.45	&	$-$0.83	&	1.85	\\
18133412$-$3148218	&	J181334.12$-$314821.82	&	4175	&	1.25	&	$-$0.85	&	1.65	\\
18133417$-$3149571	&	J181334.17$-$314957.21	&	4325	&	1.15	&	$-$0.88	&	2.10	\\
18133519$-$3149272	&	J181335.20$-$314927.26	&	4375	&	1.65	&	$-$0.90	&	2.00	\\
18133584$-$3149447	&	\nodata	&	4250	&	1.45	&	$-$0.94	&	2.10	\\
18133610$-$3148112	&	J181336.10$-$314811.20	&	4250	&	1.20	&	$-$0.86	&	2.10	\\
18133620$-$3149040	&	J181336.20$-$314904.09	&	4250	&	1.15	&	$-$0.94	&	2.00	\\
18133678$-$3150244	&	J181336.78$-$315024.51	&	4225	&	1.45	&	$-$0.87	&	2.00	\\
18133747$-$3149410	&	J181337.46$-$314941.16	&	4525	&	1.75	&	$-$0.80	&	1.90	\\
18133789$-$3147295	&	J181337.89$-$314729.53	&	4100	&	1.00	&	$-$0.92	&	2.00	\\
18133813$-$3145274	&	J181338.13$-$314527.31	&	4350	&	1.25	&	$-$0.75	&	2.10	\\
18133819$-$3149224	&	J181338.19$-$314922.45	&	4150	&	1.25	&	$-$0.90	&	1.75	\\
18134051$-$3150295	&	J181340.52$-$315029.57	&	4200	&	1.40	&	$-$0.86	&	1.85	\\
18134089$-$3149095	&	J181340.89$-$314909.61	&	4100	&	1.15	&	$-$0.86	&	1.75	\\
18134136$-$3150001	&	J181341.37$-$315000.19	&	4400	&	1.65	&	$-$0.88	&	1.85	\\
18134151$-$3148556	&	J181341.50$-$314855.52	&	4225	&	1.25	&	$-$0.79	&	1.85	\\
18134300$-$3150280	&	J181343.00$-$315028.01	&	4200	&	1.20	&	$-$0.87	&	1.95	\\
\enddata

\end{deluxetable}

%% file: tab5.tex
\tablenum{5}
\tablecolumns{14}
\tablewidth{0pt}
                   
\begin{deluxetable}{cccccccccccccc}
\rotate
\tabletypesize{\tiny}
\tablecaption{M2FS RGB Cluster Member Abundance Ratios and Uncertainties: 
Oxygen to Calcium}
\tablehead{
\colhead{Star Name}     &
\colhead{Star Name}      &
\colhead{[O/Fe]}	&
\colhead{$\Delta$[O/Fe]}      &
\colhead{[Na/Fe]}      &
\colhead{$\Delta$[Na/Fe]}      &
\colhead{[Mg/Fe]}      &
\colhead{$\Delta$[Mg/Fe]}      &
\colhead{[Al/Fe]}      &
\colhead{$\Delta$[Al/Fe]}      &
\colhead{[Si/Fe]}      &
\colhead{$\Delta$[Si/Fe]}      &
\colhead{[Ca/Fe]}      &
\colhead{$\Delta$[Ca/Fe]}      \\
\colhead{(2MASS)}	&
\colhead{(VVV)}      &
\colhead{(dex)}      &
\colhead{(dex)}      &
\colhead{(dex)}      &
\colhead{(dex)}      &
\colhead{(dex)}      &
\colhead{(dex)}      &
\colhead{(dex)}      &
\colhead{(dex)}      &
\colhead{(dex)}      &
\colhead{(dex)}      &
\colhead{(dex)}      &
\colhead{(dex)}      
} 

\startdata
18133083$-$3148103	&	J181330.83$-$314810.39	&	0.74	&	0.10	&	$-$0.03	&	0.05	&	0.43	&	0.06	&	0.42	&	0.08	&	0.39	&	0.11	&	0.07	&	0.05	\\
18133222$-$3147158	&	J181332.22$-$314715.86	&	0.58	&	0.10	&	$-$0.10	&	0.05	&	0.54	&	0.05	&	0.40	&	0.07	&	0.28	&	0.10	&	0.30	&	0.07	\\
18133324$-$3150194	&	J181333.24$-$315019.50	&	0.46	&	0.10	&	0.34	&	0.06	&	0.38	&	0.05	&	0.58	&	0.09	&	0.38	&	0.10	&	0.38	&	0.05	\\
18133412$-$3148218	&	J181334.12$-$314821.82	&	0.02	&	0.10	&	0.40	&	0.06	&	0.34	&	0.06	&	0.58	&	0.10	&	0.32	&	0.09	&	0.23	&	0.06	\\
18133417$-$3149571	&	J181334.17$-$314957.21	&	0.18	&	0.10	&	0.25	&	0.07	&	0.33	&	0.06	&	0.56	&	0.08	&	0.15	&	0.13	&	0.15	&	0.06	\\
18133519$-$3149272	&	J181335.20$-$314927.26	&	0.83	&	0.10	&	0.09	&	0.05	&	0.36	&	0.06	&	0.55	&	0.07	&	0.37	&	0.10	&	0.25	&	0.06	\\
18133584$-$3149447	&	\nodata	&	\nodata	&	\nodata	&	0.22	&	0.08	&	0.31	&	0.05	&	0.41	&	0.08	&	0.46	&	0.14	&	0.21	&	0.07	\\
18133610$-$3148112	&	J181336.10$-$314811.20	&	0.19	&	0.10	&	0.63	&	0.05	&	0.35	&	0.06	&	0.84	&	0.10	&	0.32	&	0.11	&	0.36	&	0.06	\\
\enddata

\end{deluxetable}

%% file: tab6.tex
\tablenum{6}
\tablecolumns{14}
\tablewidth{0pt}

\begin{deluxetable}{cccccccccccccc}
\rotate
\tabletypesize{\tiny}
\tablecaption{M2FS RGB Cluster Member Abundance Ratios and Uncertainties: 
Chromium to Europium}
\tablehead{
\colhead{Star Name}     &
\colhead{Star Name}      &
\colhead{[Cr/Fe]}	&
\colhead{$\Delta$[Cr/Fe]}      &
\colhead{[Fe I/H]}      &
\colhead{$\Delta$[Fe I/H]}      &
\colhead{[Fe II/H]}      &
\colhead{$\Delta$[Fe II/H]}      &
\colhead{[Ni/Fe]}      &
\colhead{$\Delta$[Ni/Fe]}      &
\colhead{[La II/Fe]}      &
\colhead{$\Delta$[La II/Fe]}      &
\colhead{[Eu II/Fe]}      &
\colhead{$\Delta$[Eu II/Fe]}	\\
\colhead{(2MASS)}	&
\colhead{(VVV)}      &
\colhead{(dex)}      &
\colhead{(dex)}      &
\colhead{(dex)}      &
\colhead{(dex)}      &
\colhead{(dex)}      &
\colhead{(dex)}      &
\colhead{(dex)}      &
\colhead{(dex)}      &
\colhead{(dex)}      &
\colhead{(dex)}      &
\colhead{(dex)}      &
\colhead{(dex)}      
} 

\startdata
18133083$-$3148103	&	J181330.83$-$314810.39	&	$-$0.18	&	0.10	&	$-$0.94	&	0.10	&	$-$0.88	&	0.11	&	$-$0.08	&	0.08	&	0.18	&	0.11	&	0.37	&	0.11	\\
18133222$-$3147158	&	J181332.22$-$314715.86	&	$-$0.12	&	0.07	&	$-$0.97	&	0.10	&	$-$0.95	&	0.12	&	$-$0.14	&	0.06	&	0.28	&	0.11	&	0.41	&	0.09	\\
18133324$-$3150194	&	J181333.24$-$315019.50	&	0.09	&	0.11	&	$-$0.81	&	0.10	&	$-$0.84	&	0.13	&	$-$0.06	&	0.06	&	0.48	&	0.11	&	0.53	&	0.10	\\
18133412$-$3148218	&	J181334.12$-$314821.82	&	$-$0.15	&	0.08	&	$-$0.85	&	0.10	&	$-$0.85	&	0.12	&	0.00	&	0.07	&	0.43	&	0.11	&	0.53	&	0.09	\\
18133417$-$3149571	&	J181334.17$-$314957.21	&	$-$0.03	&	0.09	&	$-$0.88	&	0.10	&	$-$0.87	&	0.12	&	$-$0.14	&	0.06	&	0.29	&	0.11	&	0.43	&	0.12	\\
18133519$-$3149272	&	J181335.20$-$314927.26	&	0.21	&	0.07	&	$-$0.88	&	0.10	&	$-$0.92	&	0.11	&	$-$0.06	&	0.06	&	0.52	&	0.12	&	0.55	&	0.09	\\
18133584$-$3149447	&	\nodata	&	0.01	&	0.07	&	$-$0.94	&	0.10	&	$-$0.94	&	0.12	&	$-$0.10	&	0.05	&	0.36	&	0.10	&	\nodata	&	\nodata	\\
18133610$-$3148112	&	J181336.10$-$314811.20	&	0.30	&	0.09	&	$-$0.86	&	0.10	&	$-$0.85	&	0.13	&	$-$0.12	&	0.07	&	0.68	&	0.12	&	0.51	&	0.09	\\
\enddata

\end{deluxetable}

%% file: tab7.tex
\tablenum{7}
\tablecolumns{9}
\tablewidth{0pt}

\begin{deluxetable}{ccccccccc}
\rotate
\tabletypesize{\tiny}
\tablecaption{FLAMES RGB and HB Calcium Triplet Metallicity Data for Cluster 
Members}
\tablehead{
\colhead{Star Name}     &
\colhead{Star Name}      &
\colhead{Star Name}	&
\colhead{EW$_{\rm 8542}$}      &
\colhead{EW$_{\rm 8662}$}      &
\colhead{$\Sigma$EW}      &
\colhead{W$^{\prime}$}      &
\colhead{[Fe/H]}      &
\colhead{$\Delta$[Fe/H]}      \\
\colhead{(2MASS)}	&
\colhead{(VVV)}      &
\colhead{(VLT)}      &
\colhead{(\AA)}      &
\colhead{(\AA)}      &
\colhead{(\AA)}      &
\colhead{(\AA)}      &
\colhead{(dex)}      &
\colhead{(dex)}      
} 

\startdata
18131138$-$3148128	&	J181311.39$-$314812.91	&	3002910	&	4.19	&	3.17	&	7.36	&	5.15	&	$-$0.70	&	0.01	\\
18132157$-$3151036	&	J181321.57$-$315103.68	&	3001489	&	3.43	&	2.71	&	6.14	&	5.22	&	$-$0.63	&	0.27	\\
18132381$-$3151202	&	J181323.81$-$315120.31	&	3001323	&	3.74	&	2.92	&	6.66	&	5.00	&	$-$0.82	&	0.08	\\
18132416$-$3151095	&	\nodata	&	3001259	&	3.58	&	2.60	&	6.17	&	5.31	&	$-$0.55	&	0.38	\\
18132703$-$3150191	&	J181327.03$-$315019.17	&	3000849	&	3.30	&	2.52	&	5.82	&	4.92	&	$-$0.89	&	0.04	\\
18132719$-$3155450	&	J181327.19$-$315545.14	&	3003469	&	3.81	&	2.89	&	6.70	&	4.58	&	$-$1.12	&	0.50	\\
18132987$-$3147575	&	J181329.88$-$314757.53	&	2000503	&	3.11	&	2.42	&	5.53	&	4.82	&	$-$0.96	&	0.11	\\
18132999$-$3150248	&	J181329.98$-$315024.90	&	2000789	&	2.93	&	2.14	&	5.07	&	4.65	&	$-$0.92	&	0.03	\\
\enddata

\end{deluxetable}

%% file: tab8.tex
\tablenum{8}
\tablecolumns{2}
\tablewidth{0pt}

\begin{deluxetable}{ll}
\tablecaption{Literature References}
\tablehead{
\colhead{Population}      &
\colhead{Reference}
} 

\startdata
HP-1	&	\citet{Barbuy06}	\\
HP-1	&	\citet{Barbuy16}	\\
NGC 6266	&	\citet{Yong14}	\\
NGC 6273	&	\citet{Johnson15a}	\\
NGC 6273	&	\citet{Johnson17a}	\\
NGC 6287	&	\citet{Lee02}	\\
NGC 6293	&	\citet{Lee02}	\\
NGC 6342	&	\citet{Johnson16}	\\
NGC 6342	&	\citet{Origlia05a}	\\
NGC 6352	&	\citet{Feltzing09}	\\
NGC 6362	&	\citet{Massari17}	\\
NGC 6366	&	\citet{Johnson16}	\\
NGC 6388	&	\citet{Carretta07a}	\\
NGC 6440	&	\citet{Munoz17}	\\
NGC 6440	&	\citet{Origlia08}	\\
NGC 6441	&	\citet{Gratton06}	\\
NGC 6441	&	\citet{Gratton07}	\\
NGC 6441	&	\citet{Origlia08}	\\
NGC 6522	&	\citet{Barbuy14}	\\
NGC 6522	&	\citet{Ness14}	\\
NGC 6528	&	\citet{Carretta01}	\\
NGC 6528	&	\citet{Origlia05a}	\\
NGC 6539	&	\citet{Origlia05b}	\\
NGC 6541	&	\citet{Lee02}	\\
NGC 6553	&	\citet{AlvesBrito06}	\\
NGC 6553	&	\citet{Johnson14}	\\
NGC 6553	&	\citet{Tang17}	\\
NGC 6558	&	\citet{Barbuy07}	\\
NGC 6624	&	\citet{Valenti11}	\\
NGC 6626	&	\citet{Villanova17}	\\
NGC 6681	&	\citet{OMalley17}	\\
NGC 6723	&	\citet{Gratton15}	\\
NGC 6723	&	\citet{Rojas16}	\\
Terzan 1	&	\citet{Valenti15}	\\
Terzan 4	&	\citet{Origlia04}	\\
UKS1	&	\citet{Origlia05b}	\\
Bulge Field	&	\citet{AlvesBrito10}	\\
Bulge Field	&	\citet{Gonzalez11}	\\
Bulge Field     &       \citet{Johnson12}       \\
Bulge Field	&	\citet{Bensby13}	\\
Bulge Field     &       \citet{GarciaPerez13}        \\
Bulge Field     &       \citet{Howes14}        \\
Bulge Field	&	\citet{Johnson14}	\\
Bulge Field     &       \citet{Howes15}        \\
Bulge Field	&	\citet{VanDerSwaelmen16} 	\\
\enddata

\end{deluxetable}